\documentclass[a4paper,fleqn,usenatbib]{mnras}
\usepackage{newtxtext,newtxmath}
\usepackage[T1]{fontenc}
\usepackage{ae,aecompl}
\usepackage{graphicx}	
\usepackage{subfig}
\usepackage{amsmath}	
\usepackage{amssymb}	
\title[Particle acceleration in MHD reconnection II]{Reconnection and particle acceleration in interacting flux ropes -- II. 3D effects on test particles in magnetically dominated plasmas}
\author[B. Ripperda et al.]{
B. Ripperda,$^{1}$\thanks{E-mail: bart.ripperda@kuleuven.be}
O. Porth$^{2}$,
C. Xia$^{1}$
and R. Keppens$^{1}$ 
\\
$^{1}$Centre for mathematical Plasma Astrophysics, Department of Mathematics, KU Leuven, Celestijnenlaan 200B, B-3001 Leuven, Belgium\\
$^{2}$Institut fur Theoretische Physik, Max-von-Laue-Str. 1, D-60438 Frankfurt, Germany\\
}
\date{Accepted XXX. Received YYY; in original form ZZZ}
\pubyear{2016}
\begin{document}
\label{firstpage}
\pagerange{\pageref{firstpage}--\pageref{lastpage}}
\maketitle
\begin{abstract}
We analyze particle acceleration in explosive reconnection events in magnetically dominated proton-electron plasmas. Reconnection is driven by large-scale magnetic stresses in interacting current-carrying flux tubes. Our model relies on development of current-driven instabilities on macroscopic scales. These tilt-kink instabilities develop in an initially force-free equilibrium of repelling current channels. Using MHD methods we study a 3D model of repelling and interacting flux tubes in which we simultaneously evolve test particles, guided by electromagnetic fields obtained from MHD. We identify two stages of particle acceleration; Initially particles accelerate in the current channels, after which the flux ropes start tilting and kinking and particles accelerate due to reconnection processes in the plasma. The explosive stage of reconnection produces non-thermal energy distributions with slopes that depend on plasma resistivity and the initial particle velocity. We also discuss the influence of the length of the flux ropes on particle acceleration and energy distributions. This study extends previous 2.5D results to 3D setups, providing all ingredients needed to model realistic scenarios like solar flares, black hole flares and particle acceleration in pulsar wind nebulae: formation of strong resistive electric fields, explosive reconnection and non-thermal particle distributions. By assuming initial energy equipartition between electrons and protons, applying low resistivity in accordance with solar corona conditions and limiting the flux rope length to a fraction of a solar radius we obtain realistic energy distributions for solar flares with non-thermal power law tails and maximum electron energies up to 11 MeV and maximum proton energies up to 1 GeV.
\end{abstract}
\section{Introduction}
Magnetic reconnection causes a magnetic field to rapidly and violently rearrange its topology. This topological change affects plasma energetics and is one of the processes controlling energy exchange between different plasma system constituents (\citealt{Kulsrud}; \citealt{Priest}). The main process of interest here is the conversion of energy available in the magnetic field into non-thermal particle distributions in magnetically dominated plasmas (low plasma-$\beta$). This phenomenon causes violent energy releases in a wide range of astrophysical events, including various kinds of flares and bursts of high-energy (UV, X-ray and gamma-ray) (\citealt{Uzdensky}). Solar flares are the most prominent and well studied classical examples of reconnection (\citealt{Masuda}; \citealt{Krucker}). Observations reveal that $10 \% - 50\%$ of magnetic energy is converted into energetic charged particles (\citealt{Hudson}) and that particles develop a power-law energy distribution containing energy of the same order as the converted magnetic energy (\citealt{Krucker}; \citealt{Oka}). In some observations of solar flares the emission has no distinguishable thermal part and almost all electrons are accelerated to non-thermal energies (\citealt{Krucker}; \citealt{Krucker3}). Electrons in solar flares can reach energies up to 5 -- 50 MeV, while protons gain energies up to several GeV (\citealt{Evenson}, \citealt{Aschwanden}). Reconnection has also been proposed as a mechanism for powerful flares and particle acceleration in more extreme settings like pulsar wind nebulae (\citealt{Michel}; \citealt{Lyubarsky}; \citealt{Porth3}), gamma-ray bursts (\citealt{Mckinney}), magnetospheres of magnetars (\citealt{Lyutikov}; \citealt{Lyutikov2}; \citealt{Meng}; \citealt{Elenbaas}) and in coronae and jets of accreting black holes and active galactic nuclei (\citealt{Goodman}; \citealt{Gouveia}; \citealt{Wilkins}). 

Flares from astrophysical objects require energy from macroscropic scales to be transferred to the microscopic scales on which particles are accelerated. The change of topology of the magnetic field configuration on large scales is well described by magnetohydrodynamics (MHD) and the dissipation at small scales is described by kinetic (particle) theory. The energetics of the plasma can be split into the part relevant at the fluid level plus non-thermal particle distributions. The MHD approach covers the overall scales and energetics of the system but does not give any information on particle dynamics. Kinetic approaches fully describe the microscopic scale, but are too costly to cover full astrophysical systems. In this work, we treat electrons and ions as test particles embedded in a thermal (MHD) plasma. Particle acceleration associated with reconnection and shocks in magnetically dominated Newtonian plasmas in the solar corona is studied extensively with test particle approaches (\citealt{Rosdahl}, \citealt{Gordovskyy2}, \citealt{Gordovskyy3}, \citealt{Gordovskyy4}, \citealt{Gordovskyy}, \citealt{Zhou}, \citealt{Pinto}, \citealt{Zhou2}, \citealt{Ripperda}, \citealt{Threlfall}) and even in relativistic plasmas in the context of pulsar wind nebulae (\citealt{Porth3}, \citealt{Kirk}). The test particles are guided by MHD fields without giving feedback to these fields. 

To initiate reconnection in the MHD plasma, we perturb an equilibrium of two adjacent, anti-parallel and repelling current channels (\citealt{Richard}, \citealt{Keppens}, \citealt{Ripperda}). Translation and rotation of the currents cause a plasma disruption. \cite{Ripperda} showed that reconnection occurring due to this tilt instability is an efficient source of highly energetic particles in 2.5D settings. In 3D configurations, the kink instability interacts with the tilt instability, redistributing the poloidal magnetic field. Current channels undergoing a tilt or tilt-kink instability typically show a growth phase on Alfv\'enic timescales in which kinetic energy grows exponentially. This energy is expected to be released and transferred to charged particles via magnetic reconnection. In the stellar corona context, the kink instability is one of the well-explored routes to initiate flares. Accessing a kink instability via anti-parallel, repelling flux ropes is intimately connected to the coalescence instability of two attracting and merging flux ropes. The current filaments eventually attracting or repelling typically form as a result of turbulent plasma processes or instabilities as the tearing mode. Both the tilt and coalescence instability have been studied extensively in 2D configurations (see e.g., \citealt{Longcope}; \citealt{Strauss}; \citealt{Marliani}; \citealt{Ng}). Here we propose that such anti-parallel currents may also form in stellar coronae where they play a role in magnetic island interactions and can cause particle acceleration. The evolution that is observed shows that repelling currents lead to sudden release of magnetic energy and current dissipation. The repelling islands typically cause localized reconnection, thin current sheet development and strongly curved magnetic fields. On a large scale the current interaction can lead to rapid transfer of significant amounts of energy from the currents to the particles. Our model is representative for the top parts of adjacent flux ropes as seen in extreme ultraviolet observations of the solar corona. If two of these loops develop antiparallel currents, the tilt instability route to reconnection is accessible (\citealt{Keppens}). This model was studied in 2D by \cite{Richard} and extended to 3D by \cite{Keppens} and \cite{Ripperda}.
Here we investigate the effect of an additional kink instability on particle acceleration as well as the effects of boundary conditions, initial conditions and resistivity models on particle dynamics during the full 3D MHD evolution. We use high-resolution MHD results of \cite{Ripperda} for the evolution of the repelling current channels. For test particle simulations we make use of the latest addition to the MPI-AMRVAC code (\citealt{Keppensporth}) to dynamically evolve test particle populations during MHD evolution. We apply the guiding centre approximation in which particle gyration is neglected and only the particle velocity parallel to the magnetic field is evolved. This approach is valid in typical non-relativistic, magnetically dominated plasmas where the particle energy density is less than the energy density of the underlying MHD fluid. The numerical methods employed are described in detail in Section~\ref{sec:GCA}. In Section~\ref{sec:2Dresults} 2.5D test particle simulations are discussed, in Section~\ref{sec:3Dresults} we discuss 3D simulations and the effect of the kink instability on test particle acceleration and energetics. We compare the results of the guiding centre approximation to solutions obtained from the full particle equation of motion in Appendix \ref{sect:appgyration}.

\section{Numerical setup}
\label{sec:GCA}
For the MHD background in which test particles are evolved we use a simulation as described in \cite{Ripperda}. Two parallel, adjacent, repelling current channels are initiated in a region $[-3L,3L] \times [-3L,3L] \times [-3L,3L]$ in Cartesian coordinates ($x,y,z$) with the depth of the current channel in the $z$-direction orthogonal to the plane. The equilibrium is described by the initial conditions for the flux function $\psi_0(x,y)$ for both 2.5D and 3D setups
\begin{equation}
\psi_0(x,y) = \begin{cases} \frac{2}{j_0^1 J_0 (j_0^1)}J_1(j_0^1 r) \cos(\theta) & \mbox{for } r < 1 \\ (r-\frac{1}{r}) \cos(\theta) & \mbox{for } r \geq 1, \end{cases}
\label{eq:fluxinitial}
\end{equation}
with $J_1$ is the Bessel function of the first kind and $j_0^1 \approx 3.831706$ is the first root of $J_1$. The magnetic field is obtained as $\mathbf{B} = \nabla \psi_0 \times \mathbf{\hat{z}} + B_z \mathbf{\hat{z}}$, resulting in a current distribution $J_z = (\nabla \times \mathbf{B})_z = \nabla^2 \psi_0$. In one half of the unit circle $J_z < 0$ and in the other half $J_z > 0$ and initially there are no currents in the region $r \geq 1$. An ideal MHD equilibrium is established by postulating a force-free magnetic field with spatially varying, vertical component $B_z(x,y)$ and a uniform plasma pressure $p_0$ such that the Lorentz force $\mathbf{J} \times \mathbf{B} = \nabla p = 0$, here
\begin{equation}
B_{z}(x,y) = \begin{cases} (j_0^1)(\psi_0(x,y)) & \mbox{for} r < 1. \\ 0 & \mbox{for } r \geq 1. \end{cases}
\label{eq:eqB}
\end{equation}
In 2.5D configurations a uniform resolution of $2400^2$ is used and in 3D setups we use an effective resolution of $300^3$ with one level of mesh refinement. $L = 10$ Mm is chosen as a typical unit of length for the astrophysical systems under consideration. The dimensionless density $\rho$ is equal to unity initially and the ratio of specific heats is $\Gamma=5/3$. We fix a constant pressure $p_0 = 0.01/\Gamma$ such that we reach a low plasma-$\beta = 0.04$ in an initially force-free equilibrium. The normalization used implies the sound speed outside the current channels as the unit of speed, the radius of the double current channel as the unit of length and the density to fix the unit of mass. Magnetic units where $\mu_0=1$ are employed. MHD fields and particles are evolved for the typical time $10 t_S = 10 L/c_S \approx 852.6$ seconds, with $c_S$ the sound speed outside the current channels. We apply a uniform resistivity $\eta_{MHD}=10^{-4}$. The low-$\beta$ and force-free conditions are in accordance with astrophysical systems which are magnetically dominated, such as magnetospheres of black holes and pulsars and the solar corona. The boundary conditions for the MHD evolution in the $(x,y)$-plane imply a zero gradient. The boundaries in the $z$-direction are periodic in 3D configurations, whereas in 2.5D the $z$-direction is invariant. The initially force-free equilibrium is unstable to a tilt instability (\citealt{Richard}) and in a 3D configuration also to an additional kink instability bending the field lines with respect to the vertical direction (\citealt{Keppens}; \citealt{Ripperda}). Once the instabilities develop and the physics become naturally nonlinear, it allows for fast reconnection of the field lines. 

From solving the set of resistive, compressible MHD equations we obtain the magnetic and electric fields $\mathbf{B}$ and $\mathbf{E}$ and the total current density $\mathbf{J} = \nabla \times \mathbf{B}$. The MHD data are scaled to CGS units before being used in the test particle calculations. To analyze the energetics and acceleration of electrons and protons in the MHD simulations we follow the orbits of test particles in the MHD flow, similar to the approach of \cite{Porth3} and \cite{Ripperda}. In 3D simulations particles are injected uniformly in the domain. To analyze the behavior of particles in the reconnection zones specifically, several runs are performed with a fraction of $0.99$ of the ensemble uniformly distributed in space in a rectangular block, encapsulating the two (displaced) current channels and the areas with the largest current density, $x \in [-1L,1L]$, $y \in [-2L,2L]$, $z \in [-3L,3L]$. The other fraction of $0.01$ of the ensemble is uniformly distributed over the full domain $x \in [-3L,3L]$, $y \in [-3L,3L]$, $z \in [-3L,3L]$, including the surrounding background. In 2.5D simulations, with translational invariance in the $z$-direction for MHD evolution, the particles are distributed in the $(x,y)$-plane in accordance with the 3D simulations, at $z = 0$. In case I3De (see Table~\ref{tab:example_table}) the current channels are twice as long as in the reference cases, meaning that the particles are distributed over the domain $x \in [-3L,3L]$, $y \in [-3L,3L]$, $z \in [-6L,6L]$ and the MHD resolution is $300 \times 300 \times 600$.

Typical parameters for low plasma-$\beta$ plasma in coronal loops are, for magnetic field magnitude $B=0.03 T$, temperature $T = 10^6 K$, number density $n = 10^{16}$ $m^{-3}$ and plasma-$\beta = 0.0004$ (\citealt{Goedbloed}). Particles are injected from a Maxwell-Boltzmann velocity distribution in accordance with solar corona conditions
\begin{equation}
f(v) = N\left(\frac{2v}{v^{2}_{th}}\right) \exp\left(\frac{-v^2}{v^{2}_{th}}\right).
\label{eq:Maxwellian}
\end{equation}
with thermal speed $v_{th,p} = \sqrt{(2 k_B T \rho_0/m_{p} p_0)} \sim 10^7 m/s$ for protons in a fluid with temperature $T = 10^6 K$, the proton rest mass $m_p = 1.6726 \cdot 10^{-24} g$, dimensionless pressure $p_0$ and fluid density $\rho_0$. For electrons we either assume energy equipartition, meaning they have a thermal speed $v_{th,e} = \sqrt{m_p/m_e} \times v_{th,p} \sim 10^9 m/s$, or we assume that both electrons and protons initially have the typical thermal speed $v_{th,e} = v_{th,p} = \sqrt{(2 k_B T \rho_0/m_{p} p_0)}$. Both resulting in a thermal Lorentz factor of $\gamma_{th} \approx 1$. The particle gyroradius in the plasma settings we assume, $R_L = \gamma m_0 v_{\perp}/(Bq) = 10^{-3} m$ for electrons and $R_L = 4.4 \times 10^{-2} m$ for protons, is small compared to the typical size over which the MHD fields change. The particles have a uniform pitch angle distribution $\alpha \in [-\pi/2, \pi/2]$ with $\alpha = \arctan(v_{\perp}/v_{\|})$, the angle between the velocity vector of a particle and the unit vector parallel to the magnetic field. 

The particles are advanced according to the Lorentz force resulting from the MHD fields $\mathbf{E}$ and $\mathbf{B}$ (e.g. \citealt{Landau}):
\begin{equation}
\frac{d\mathbf{U}}{dt} = \frac{q}{m_0 c}\left(\mathbf{E} + \frac{\mathbf{U}\times\mathbf{B}}{c\gamma}\right),
\label{eq:lorentztens}
\end{equation}
where $\mathbf{U} = \gamma \mathbf{v}/c$ is the particles four-velocity, $c$ the speed of light in vacuum and $q/m_0$ is the charge to mass ratio. We apply the guiding centre approximation, in which the gyration of the particles is neglected, to equation (\ref{eq:lorentztens}) to obtain the relativistic guiding centre equations of motion describing the (change in) guiding centre position $\mathbf{R}$, parallel relativistic momentum $p_{\|} = m_0\gamma v_{\|}$ and relativistic magnetic moment $\mu_r = m_0 \gamma^{2} v^{2}_{\perp}/2B$ in three-space (\citealt{Vandervoort})
\begin{align*}
\frac{d\mathbf{R}}{dt} = \frac{\left(\gamma v_{\|}\right)}{\gamma}\mathbf{\hat{b}}+\frac{\mathbf{\hat{b}}}{B\left(1-\frac{E_{\perp}^{2}}{B^2}\right)} \times \Biggl\{ -\left(1-\frac{E_{\perp}^{2}}{B^2}\right)c\mathbf{E} + \Biggr. \nonumber
\end{align*}
\begin{align*}
\frac{cm_0\gamma}{q}\left(v_{\|}^{2}\left(\mathbf{\hat{b}}\cdot\nabla\right)\mathbf{\hat{b}}+v_{\|}\left(\mathbf{u_E}\cdot\nabla\right)\mathbf{\hat{b}} + v_{\|}\left(\mathbf{\hat{b}}\cdot\nabla\right)\mathbf{u_E} + \left(\mathbf{u_E}\cdot \nabla\right)\mathbf{u_E}\right) + \nonumber
\end{align*}
\begin{align}
\Biggl. \frac{\mu_r c}{\gamma q}\nabla\left[B\left(1-\frac{E_{\perp}^{2}}{B^2}\right)^{1/2}\right]  + \frac{v_{\|}E_{\|}}{c}\mathbf{u_E}  \Biggr\},
\label{eq:gcastatic1}
\end{align}
\begin{align}
\frac{d \left(m_0 \gamma v_{\|}\right)}{dt} =  m_0\gamma\mathbf{u_E}\cdot \left(v_{\|}\left(\mathbf{\hat{b}}\cdot\nabla\right)\mathbf{\hat{b}}+\left(\mathbf{u_E}\cdot\nabla\right)\mathbf{\hat{b}}\right) +\nonumber \\ 
qE_{\|} -\frac{\mu_r}{\gamma}\mathbf{\hat{b}}\cdot\nabla\left[B\left(1-\frac{E^{2}_{\perp}}{B^2}\right)^{1/2}\right],
\label{eq:gcastatic2}
\end{align}
\begin{equation}
\frac{d \left(m_0 \gamma^{*2} v^{*2}_{\perp}/2B^*\right)}{dt} = \frac{d \mu_{r}^{*}}{dt} = 0.
\label{eq:gcastatic3}
\end{equation}
Here, $\mathbf{\hat{b}}$ is the unit vector in the direction of the magnetic field and $v_{\|}$ the component of the particle velocity vector parallel to $\mathbf{\hat{b}}$. The magnitude of the electric field $\mathbf{E} = -\mathbf{v} \times \mathbf{B} + \eta_{p} \mathbf{J}$ is split as $E = \sqrt{E^{2}_{\perp} +E^{2}_{\|}}$ where the component parallel to the magnetic field, $\mathbf{E_{\|}}$, comes solely from resistive contributions $\eta_{p} \mathbf{J} \cdot \mathbf{\hat{b}}$ and is therefore also called the resistive electric field. The resistivity $\eta_p$ is either equal to the resistivity set for the MHD evolution $\eta_p = \eta_{MHD}$ or it is an anomalous resistivity $\eta_p \neq \eta_{MHD}$ that does not affect MHD fields. The drift velocity, perpendicular to $\mathbf{B}$ is written as $\mathbf{u_E} = c\mathbf{E}\times\mathbf{\mathbf{\hat{b}}}/B$ and $v^{*}_{\perp}$ is the perpendicular velocity of the particle, in the frame of reference moving at $\mathbf{u_E}$. The magnetic field in that frame is given by $B^* = B(1-E^{2}_{\perp}/B^2)^{1/2}$ up to first order. The relativistic magnetic moment $\mu_{r}^{*}$ is an adiabatic invariant and is proportional to the flux through the gyration circle, again in the frame of reference moving at $\mathbf{u_E}$. The oscillation of the Lorentz factor at the gyrofrequency is averaged out as well, giving $\gamma = \gamma^{*}(1-E^{2}_{\perp}/B^2)^{-1/2}$. We assume the MHD fields to be slowly varying compared to the particle dynamics, allowing to neglect temporal derivatives in equations (\ref{eq:gcastatic1}-\ref{eq:gcastatic3}). However, we do treat dynamic MHD evolutions, so we interpolate the MHD variables in time. In the case of the guiding centre approximation, the gyroradius of the particle, $R_L = \gamma m_0 v_{\perp}/(Bq)$, is assumed to remain smaller than the typical cell size of the MHD simulation. To confirm validity of the guiding centre approximation results of both equations (\ref{eq:lorentztens}) and (\ref{eq:gcastatic1}-\ref{eq:gcastatic3}) will be compared. For more information on the guiding centre approach used here and its validity we refer to \cite{Ripperda} and for its mathematical background to \cite{Northropbook}.

Equations (\ref{eq:lorentztens}) are advanced with a second-order symplectic Boris scheme (e.g. \citealt{Birdsall}). Each particle is advanced with an adaptive, individual time step, ensuring that a single gyration is resolved by at least 60 steps, to ensure numerical stability. Equations (\ref{eq:gcastatic1}-\ref{eq:gcastatic3}) are advanced with a fourth order Runge-Kutta scheme with adaptive time stepping. Here, the particle timestep $\delta t$ is determined based on its parallel acceleration $a = d v_{\|}/dt$ and velocity $v = \sqrt{(v_{\|})^2 + (v_{\perp})^2}$ as the minimum of $\delta r / v$ and $v / a$, where $\delta r$ is the grid step. This grid step is restricted such that a particle cannot cross more than one cell of the MHD grid in one time step. The fields $\mathbf{E}$ and $\mathbf{B}$, and for the GCA equations their spatial derivatives, are obtained at the particles position via linear interpolations in space and time between the fluid steps limited by the CFL condition. The particles gyroradius is also calculated at every timestep and compared to the typical cell size to monitor the validity of the guiding centre approximation.

In the $(x,y)$-plane we employ open boundary conditions, in which the particles leaving the physical domain are destroyed. In 2.5D simulations we limit the length of the flux ropes in the $z$-direction, which is invariant for MHD fields. A particle crossing an artificially set boundary, at $z=3 L$ or $z=-3L$ (consistent with the $z$-boundaries in 3D simulations), is destroyed. For each destroyed particle a new particle is injected at the opposite $z$-boundary with a thermal velocity from a Maxwellian distribution. This thermal bath boundary condition limits the length of the flux rope to $6L$ and counteracts particles accelerating indefinitely in the invariant $z$-direction. In 3D configurations, we have periodic boundary conditions for MHD fields, where particles leaving a $z$-boundary are periodically injected at the opposite $z$-boundary, consistent with MHD. In specific cases (see Table \ref{tab:example_table}) we employ a similar boundary condition as in 2.5D configurations where particles leaving a $z$-boundary are destroyed. For each destroyed particle a new particle is injected at the opposite boundary with a thermal velocity according to a Maxwellian. In this way a large enough ensemble of particles is retained at all times, to achieve accurate statistics. Consequently the length of the flux rope is limited to $6L$ or equivalently $\sim 0.1$ solar radii. This boundary condition realistically mimics the injection of thermal particles in a (curved) flux rope in the corona of a star. 

To go to more realistic solar corona conditions we set a particle resistivity $\eta_p$, that does not affect the MHD evolution. The factor $\eta_{p}$ appears in all terms in equations (\ref{eq:gcastatic1}-\ref{eq:gcastatic2}) as $E_{\|} = \eta_{p} \mathbf{J} \cdot \mathbf{\hat{b}} $ where the current $\mathbf{J}$ is interpolated at the particle position from MHD fields. This parameter is either set to be equal to the resistivity used for the MHD evolution $\eta_p = \eta_{MHD} = 10^{-4}$, or set smaller than the MHD resistivity as $\eta_{p} = 10^{-5} \eta_{MHD} = 10^{-9}$. Decreased resistivity avoids artificially large energies due to a large resistive electric field. The magnetic Reynolds number, describing the ratio of advective to diffusive terms in the induction equation is defined by
\begin{equation}
R_m = \mathcal{U} \mathcal{L} /\eta_D.
\label{eq:magneticreynolds}
\end{equation}
with $\mathcal{U}$ and $\mathcal{L}$ the characteristic velocity and length scale respectively and $\eta_D = \eta L c_S$ the dimensional resistivity and $c_S$ the sound speed time used as unit of velocity. In the solar corona it is typically $\mathcal{O}(10^8)$ -- $\mathcal{O}(10^{12})$ (\citealt{Hood}). Large values of the magnetic Reynolds number mean that resistive effects are restricted to thin regions with large current density. In our simulations the typical length scale is $\mathcal{L} = 6 L = 6 \cdot 10^9$ cm, the total width of the simulation box (the maximum diameter of the flux tubes). The typical velocity is the Alfv\'en speed $\mathcal{U} = V_A  = B/\sqrt{\rho} \approx 43 \cdot 10^6$ cm/s. With $\eta_p = \eta_{MHD} = 10^{-4}$ we find $R_m \approx 2 \cdot 10^5$ and for $\eta_p = 10^{-9}$ we find $R_m = 2 \cdot 10^{10}$. Simulations with different $R_m$ are compared to see the effect of the resistivity on particle acceleration, without changing the MHD results. 

In Table \ref{tab:example_table} we list all particle simulations including the dimension, the type of particles (indicated by $e^-$ for electrons and $p^+$ for protons), the number of particles $N_{tot}$, the initial spatial distribution (which is always uniform in the domain given) and the equations of motion solved for the particles (guiding centre approximation, indicated with GCA or full equations of motion indicated with Lorentz). We indicate the typical length $l$ the particles can travel in the current channels in the $z$-direction. Infinite length corresponds to periodic boundary conditions. A finite value corresponds to distance between the two opposite $z$-boundaries at which a particle is typically destroyed and a thermal particle is injected respectively. We also mention the resistivity used in the particles equations of motion $\eta_{p}$ (either $10^{-9}$ or equal to the resistivity applied for the MHD evolution $\eta_{MHD} = 10^{-4}$), and the magnetic Reynolds number $R_m$ typical for the simulation parameters. In the last two columns we show the particles maximum kinetic energy $\mathcal{E}_{kin,max}/(m_0 c^2) = \gamma_{max} -1$ and the particles maximum energy $\mathcal{E}_{max} = \gamma m_0 c^2$ in MeV for each run.
\begin{table*}
	\centering
	\caption{The simulated cases and several characteristic parameters.}
	\label{tab:example_table}
	\begin{tabular}{lcccccccccr} 
		\hline
		Run&Particle&$N_{tot}$&Spatial distribution&Equations&$l$&$\eta_p$&$R_m$&$v_{th}$&$\gamma_{max}-1$&$\mathcal{E}_{max}$ [MeV]\\
		\hline
		A2De&$e^-$&20.000&$|x| \leq 1L$; $|y| \leq 2L$; $z=0$&GCA&$6L$&$10^{-4}$&$2 \cdot 10^{5}$&$v_{th,p}$&$1 \cdot 10^2$&$52$\\
		B2Dp&$p^+$&20.000&$|x| \leq 1L$; $|y| \leq 2L$; $z=0$&GCA&$6L$&$10^{-4}$&$2 \cdot 10^{5}$&$v_{th,p}$&$2 \cdot 10^{-2}$&$96 \cdot 10^1$\\
		\hline
		A3De&$e^-$&20.000&$|x| \leq 1L$; $|y| \leq 2L$; $|z| \leq 3L$&GCA& $6L$ & $10^{-4}$&$2 \cdot 10^{5}$&$v_{th,p}$&$6 \cdot 10^{1}$&$31$\\
		B3Dp&$p^+$&20.000&$|x| \leq 1L$; $|y| \leq 2L$; $|z| \leq 3L$&GCA& $6L$ & $10^{-4}$&$2 \cdot 10^{5}$&$v_{th,p}$&$2 \cdot 10^{-2}$&$96 \cdot 10^1$ \\
		C3De&$e^-$&20.000&$|x| \leq 1L$; $|y| \leq 2L$; $|z| \leq 3L$&GCA& $\infty$ & $10^{-9}$& $2 \cdot 10^{10}$& $v_{th,p}$&$1 \cdot 10^{3}$&$51 \cdot 10^1$\\
		D3Dp&$p^+$&20.000&$|x| \leq 3L$; $|y| \leq 3L$; $|z| \leq 3L$&GCA& $\infty$ & $10^{-9}$& $2 \cdot 10^{10}$& $v_{th,p}$&$2 \cdot 10^{-2}$&$96 \cdot 10^1$ \\
		E3De&$e^-$&20.000&$|x| \leq 1L$; $|y| \leq 2L$; $|z| \leq 3L$&GCA& $6L$ & $10^{-9}$& $2 \cdot 10^{10}$& $v_{th,p}$&$5$&$3.1$ \\
		F3Dp&$p^+$&20.000&$|x| \leq 3L$; $|y| \leq 3L$; $|z| \leq 3L$&GCA& $6L$ & $10^{-9}$& $2 \cdot 10^{10}$& $v_{th,p}$&$3 \cdot 10^{-2}$&$97 \cdot 10^1$\\
		G3De&$e^-$&20.000&$|x| \leq 3L$; $|y| \leq 3L$; $|z| \leq 3L$&GCA&$6L$ & $10^{-9}$& $2 \cdot 10^{10}$&$v_{th,p}\sqrt{\frac{m_{p}}{m_{e}}}$&$7$&$4.1$ \\
		H3De&$e^-$&20.000&$|x| \leq 1L$; $|y| \leq 2L$; $|z| \leq 6L$&GCA&$12L$ & $10^{-9}$&$2 \cdot 10^{10}$&$v_{th,p}\sqrt{\frac{m_{p}}{m_{e}}}$&$2 \cdot 10^{1}$&$11$ \\
		I3De&$e^-$&200.000&$|x| \leq 1L$; $|y| \leq 2L$; $|z| \leq 3L$&GCA& $6L$ & $10^{-4}$& $2 \cdot 10^{5}$& $v_{th,p}$&$6 \cdot 10^{1}$&$31$ \\
		J3De&$e^-$&20.000&$|x| \leq 3L$; $|y| \leq 3L$; $|z| \leq 3L$&GCA& $6L$ & $10^{-9}$& $2 \cdot 10^{10}$& $v_{th,p}$&$5$&$3.1$ \\		
		K3Dp&$p^+$&20.000&$|x| \leq 3L$; $|y| \leq 3L$; $|z| \leq 3L$&Lorentz&$\infty$&$10^{-9}$&$2 \cdot 10^{10}$&$v_{th,p}$&$4 \cdot 10^{-2}$&$98 \cdot 10^1$ \\
		\hline
	\end{tabular}
			
\textbf{Note}:The leftmost column labels the various runs. The other columns quantify various initial particle parameters. The two rightmost columns indicate the (approximate) maximum kinetic energy $E_{kin,max}/(m_0 c^2) = \gamma_{max}-1$ and the (approximate) maximum energy $\gamma_{max} m_0 c^2$ in MeV  (see text for details).
\end{table*}

\section{Results in 2.5D configurations}
\label{sec:2Dresults}
\begin{figure}
\includegraphics[width=\columnwidth]{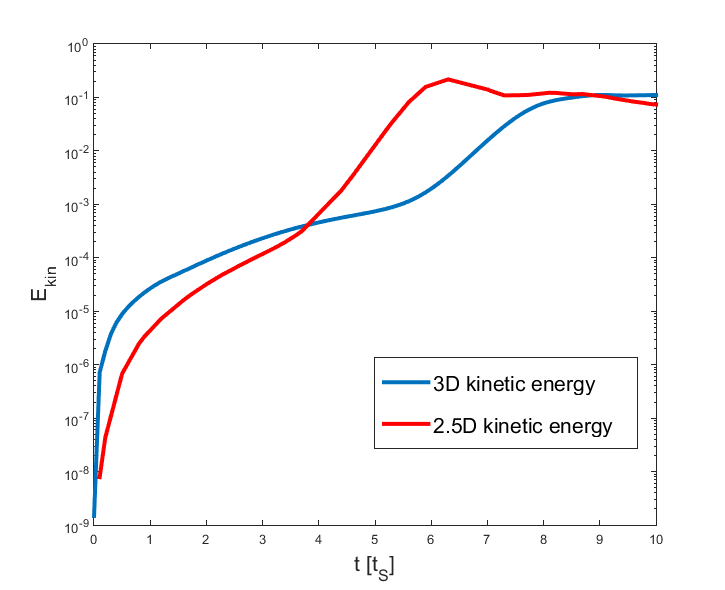}
    \caption{Kinetic energy density evolution of the MHD fluid for 2.5D with effective resolution $2400^2$ and 3D with effective resolution $300^3$, at all times integrated over a single current channel as identified by an advected tracer.}
    \label{fig:kineticenergy}
\end{figure}

\begin{figure*}
  \centering
		\subfloat{\includegraphics[width=\columnwidth, clip=true]{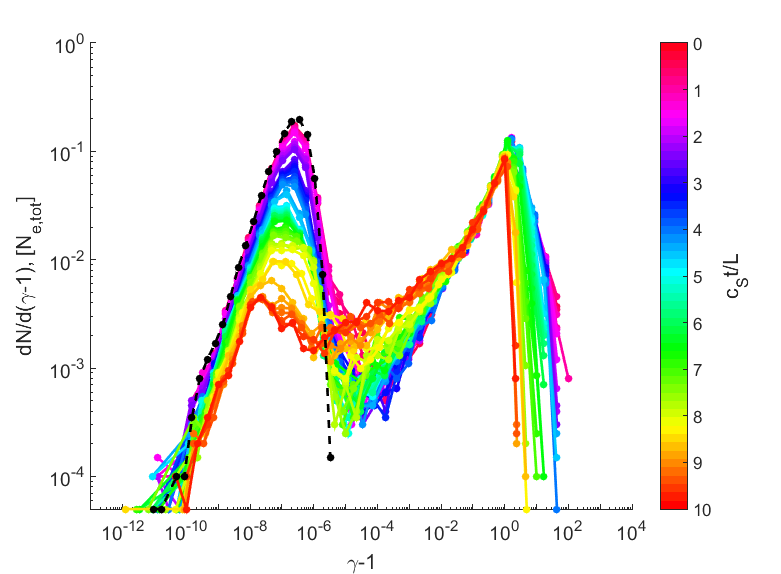}}
						\subfloat{\includegraphics[width=\columnwidth, clip=true]{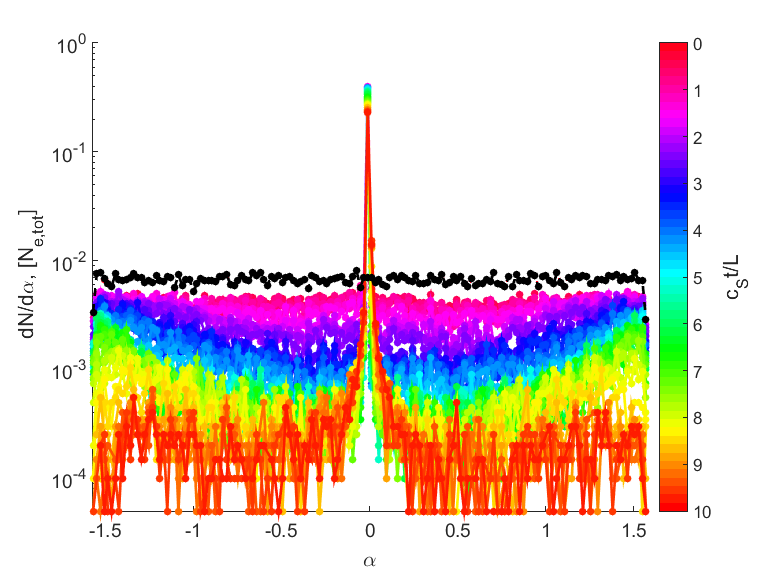}}
		\caption{2.5D Kinetic energy distribution counted by particle number (left-hand panel) and pitch angle distribution (right-hand panel) for case A2De with 20.000 electrons from $t=0$ to $t=10$, solved for the guiding centre approximation with $\eta_p = 10^{-4}$, with thermal bath applied and thermal velocity $v_{th,p}=\sqrt{2 k_B T \rho_0/(m_{p} p_0)}$. Time is measured in units of $L/c_S$, see the colour bar at the right, with $L=10 \cdot 10^6 m$ and $c_S$ the speed of sound. The initial Maxwellian is depicted with a dashed, black line, showing the thermal part of the distribution.}
\label{fig:Ekin_distributions_20000electrons_2D}
\end{figure*}

Flares are strongly transient phenomena and particles accelerate during such an event. Recently, \cite{Ripperda} combined MHD and test particle methods to investigate proton and electron acceleration in static MHD snapshots of repelling flux tubes in 2.5D. Here particle dynamics are evolved simultaneously with MHD evolution. The 2.5D results presented by \cite{Ripperda} show very hard energy distributions and an inverted power law spectrum due to indefinite acceleration in the infinitely long current channels. Here we suggest several solutions to obtain more realistic distributions with a power law spectrum and energies in accordance with observations, both in 2.5D setups and 3D setups. The perturbed equilibrium of adjacent and anti-parallel currents develops a tilt instability in which the current channels start to displace and rotate. This instability is indicated by an exponential growth phase of the kinetic energy, that is reached after $t \approx 4 t_S$ in our 2.5D simulation (see Fig.~\ref{fig:kineticenergy}).  After this phase, the non-linear regime is reached at $t \approx 6 t_S$, showing highly chaotic behavior and magnetic energy is converted into kinetic energy. In 3D both phases are delayed until $t \approx 6 t_S$ and $t \approx 8 t_S$ respectively, due to the magnetic tension caused by the kinking of the channels (see Fig.~\ref{fig:kineticenergy}). We evolve particles during the whole evolution shown in Fig. \ref{fig:kineticenergy}, however we are mainly interested in particles accelerating due to the tilt instability from $t \approx 6 t_S$ onwards. 
Particle acceleration is quantified by means of energy distributions and pitch angle distributions. Electron energy distributions associated with solar flares typically have a high-energy tail that partially can be fitted with a power law function $f(\mathcal{E}) \propto (\mathcal{E})^{-p}$ with $p \geq 1$. A longer time spent in the current channels corresponds to higher energies, and harder energy distributions. In \cite{Ripperda} it is shown that interacting flux ropes in 2.5D configurations are an efficient mechanism to accelerate particles. However, the spectra found are very hard and the power law slope is even inverted ($p < 0$) compared to what is expected based on observations, even on very short time scales $\Delta t \ll 0.1 t_S$. The maximum energies found also exceed electron energies of 5 -- 50 MeV associated with solar flares. The main cause mentioned is the 2.5D character of the setup, meaning that the current channels have an infinite length and hence, particles can accelerate indefinitely. Here the length of the flux ropes is limited to $6L$ by applying a thermal bath at $z = \pm 3L$, in accordance with the periodic boundaries in 3D simulations.
In Fig.~\ref{fig:Ekin_distributions_20000electrons_2D} we show the kinetic energy ($\mathcal{E}_{kin}/(m_0 c^2) = \gamma - 1$) distribution (left-hand panel) and the pitch angle ($\alpha = \arctan(v_{\|}/v_{\perp})$) distribution (right-hand panel) counted by particle number, for electrons in case A2De. The spectra are coloured by the MHD time $t_S$, from magenta to red, with magenta corresponding to early times and red to late times. The initial distributions are depicted by a dashed black line. Initially the electrons are distributed in the regions of the (displaced) current channels $-1L \leq x \leq 1L$; $-2L \leq y \leq 2L$ from a Maxwellian with thermal speed $v_{th,e} = v_{th,p} = \sqrt{2 k_B T \rho_0/(m_{p} p_0)}$, to improve statistics of particles in reconnection regions. We have confirmed that a simulation with an initially uniform electron distribution gives similar results, with the same maximum energy $\gamma_{max}$ but a smaller fraction of particles in the high energy tail. The limited flux rope length bounds the kinetic energy to $\gamma - 1 \lesssim 10^2$, corresponding to $\mathcal{E}_{max} \approx 50$ MeV, which is three orders of magnitude smaller than in the case with infinitely long flux ropes (\citealt{Ripperda}). Acceleration in the current channels at early times causes the slope of the spectrum to be inverted, $p < 0$. Acceleration in the direction parallel to the magnetic field is dominant over particle drifts as can be seen from the pitch angle distribution in the right panel of Fig.~\ref{fig:Ekin_distributions_20000electrons_2D} that is strongly peaked around $\alpha = 0$.
For protons (case B2Dp in Table \ref{tab:example_table}) the kinetic energy spectra look fairly similar, with two major differences; The maximum kinetic energy is limited by $\gamma - 1 \lesssim 10^{-2}$, corresponding to $E_{max} \approx 957$ MeV, due to the mass difference between electrons and protons, and less protons have left the domain through the open $x$-, and $y$-boundaries. The power law index of the high energy tail is inverted, $p < 0$. In the pitch angle spectra an asymmetry with respect to $\alpha = 0$ can be observed, due to electrons and protons accelerating in opposite directions. This observation is visible in all simulations carried out in this work. The tendency for particle to accelerate along the magnetic field lines, and hence obtain a very small pitch angle, is in agreement with the findings of \cite{Gordovskyy}. However, curvature acceleration resulting in increasing parallel velocity is neglected in their setup and particle collisions are incorporated. This asymmetry develops directly after $t=0$, when protons accelerate parallel to the magnetic field. Therefore there are more particles with a pitch angle slightly larger than zero. For electrons this asymmetry is present as well, with more particles with a pitch angle slightly smaller than zero. However, because electrons develop a larger parallel velocity than protons, the $\alpha=0$ peak is sharper for electrons and the asymmetry around $\alpha=0$ is less pronounced than for protons. For protons the $\alpha = 0$ peak is represented by a peak at $\cos(\zeta) = 1$ and for electrons at $\cos(\zeta) = -1$ if we define $\cos(\zeta) = v_{\|} / v$ with $\zeta$ the angle between the velocity vector and the magnetic field vector. 

\section{Results in 3D configurations}
\label{sec:3Dresults}
\begin{figure*}
\includegraphics[width=0.666666667\columnwidth, trim= 11.65cm 2cm 11.15cm 2cm, clip=true]{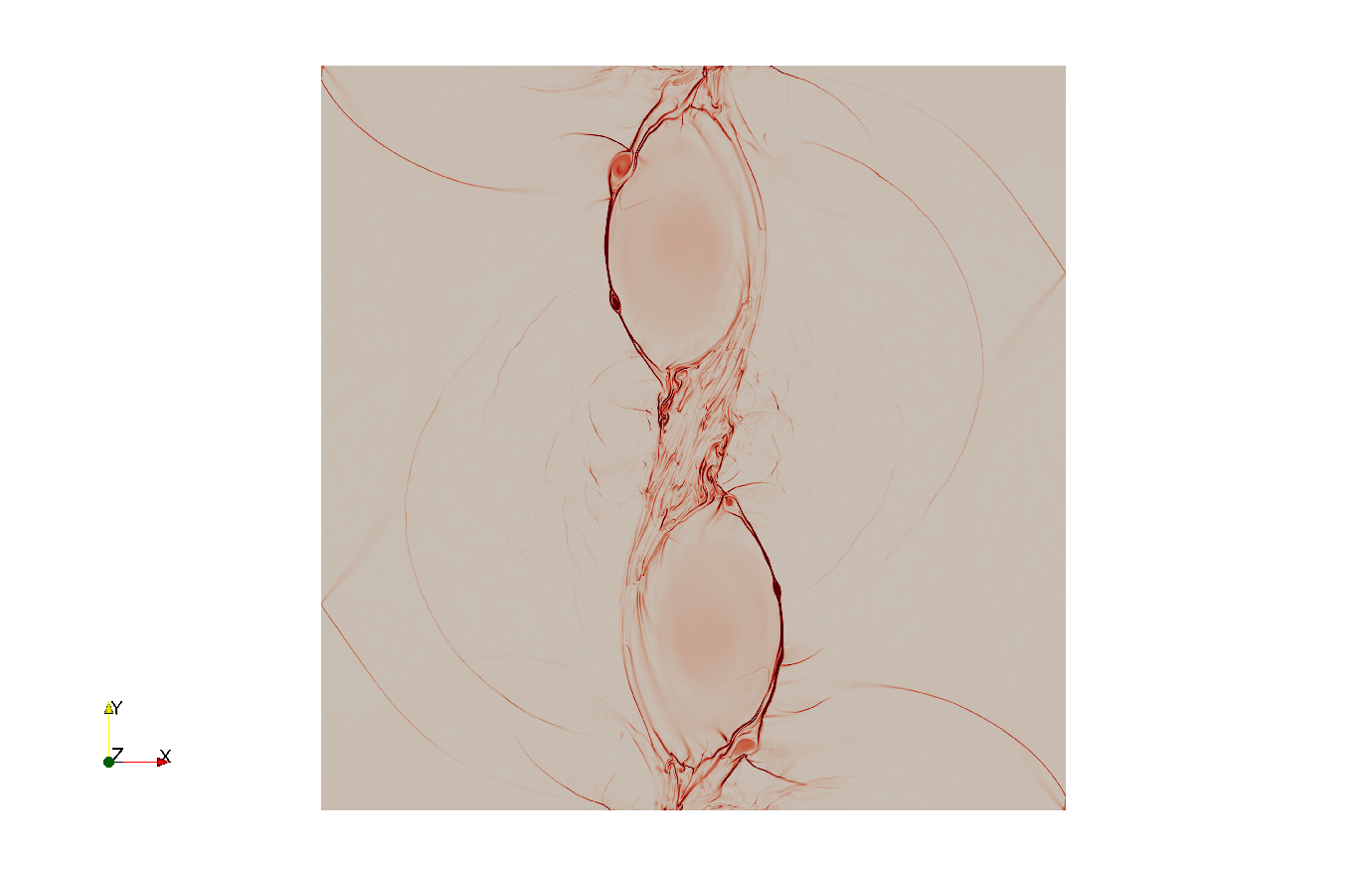}
\includegraphics[width=0.666666667\columnwidth, clip=true]{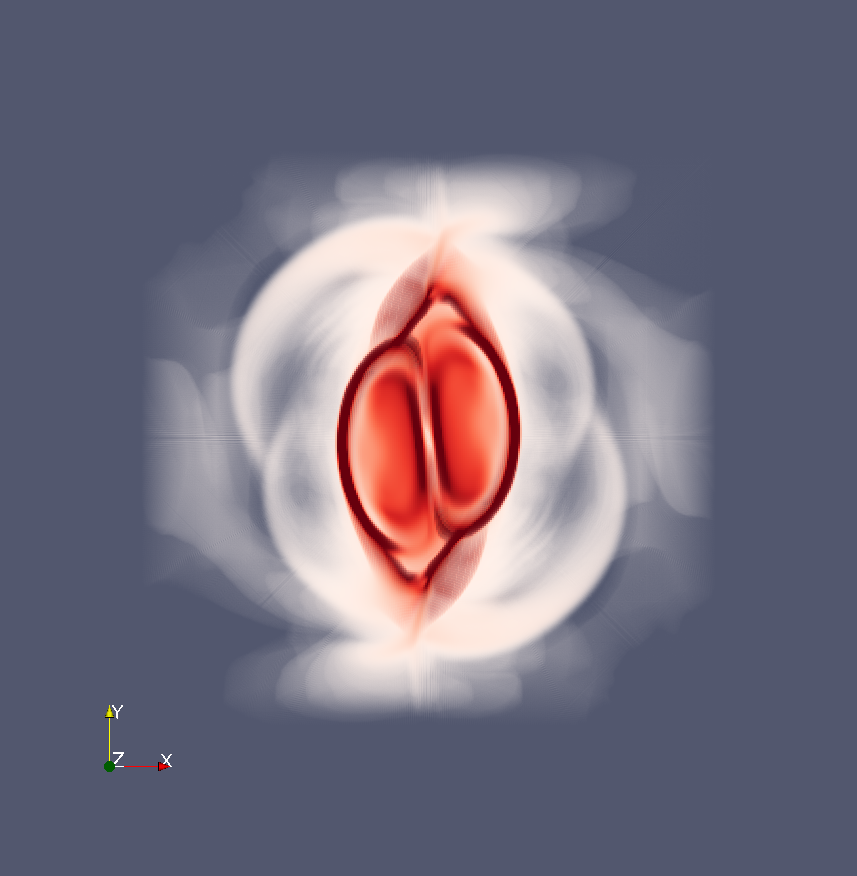}
\includegraphics[width=0.666666667\columnwidth, clip=true]{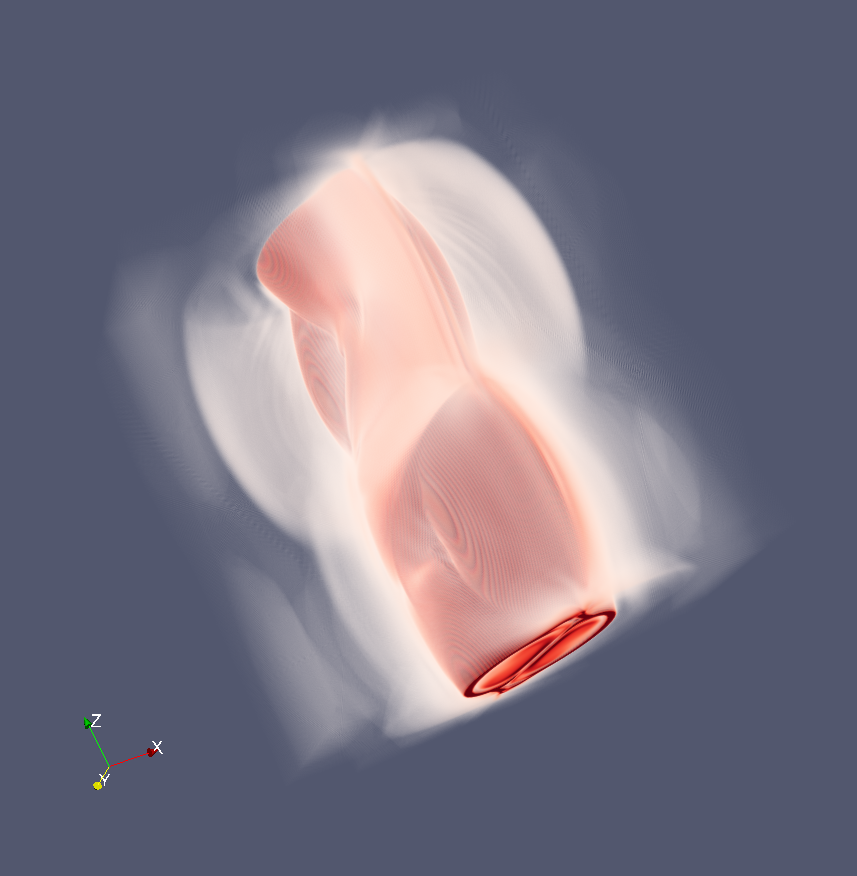}
\caption{The total current density magnitude $|\mathbf{J}|$ with a linear colour scale saturated to show values between $[0, 50]$ in 2.5D (left-hand panel) and between $[0, 10]$ 3D (middle for a top view and right-hand panel for a side view). The secondary islands are visible in the 2.5D case, whereas thin current sheets are visible both in 2.5D and 3D, indicated by a srtong current magnitude.}
\label{fig:Jtot2Dand3D}
\end{figure*}

\begin{figure*}
\includegraphics[width=\columnwidth, clip=true]{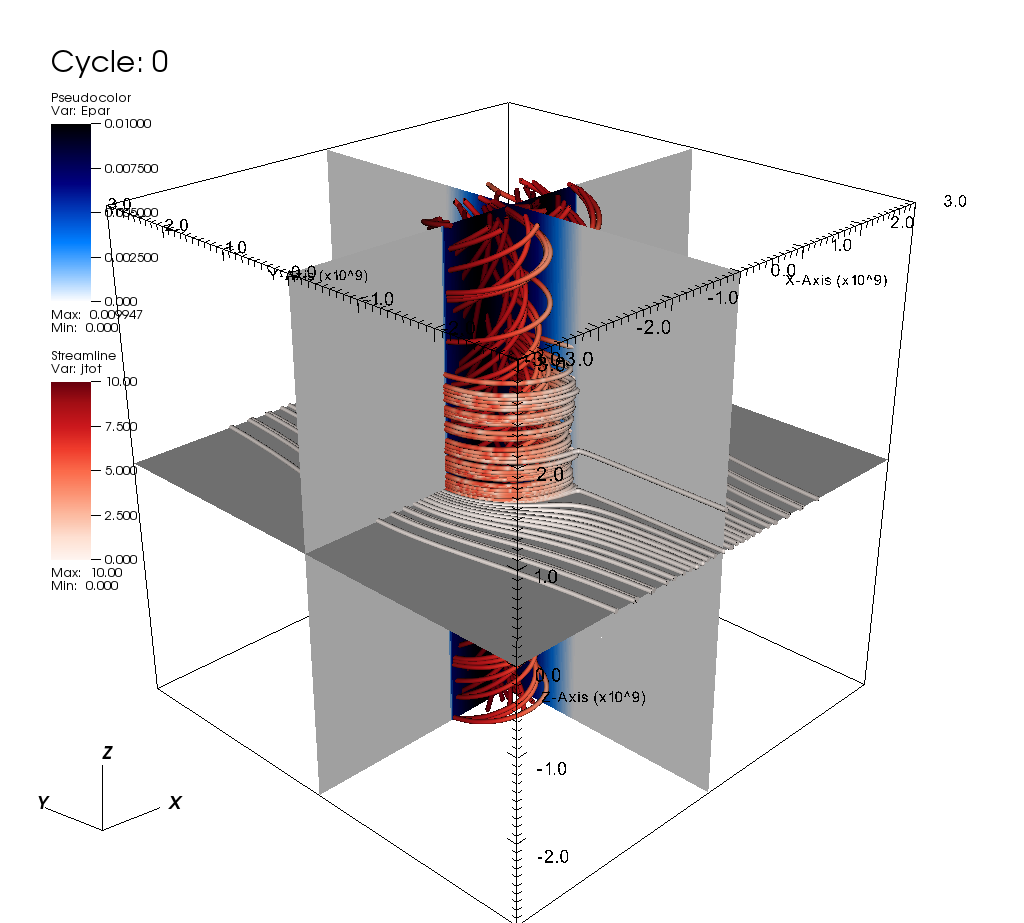}
\includegraphics[width=\columnwidth, clip=true]{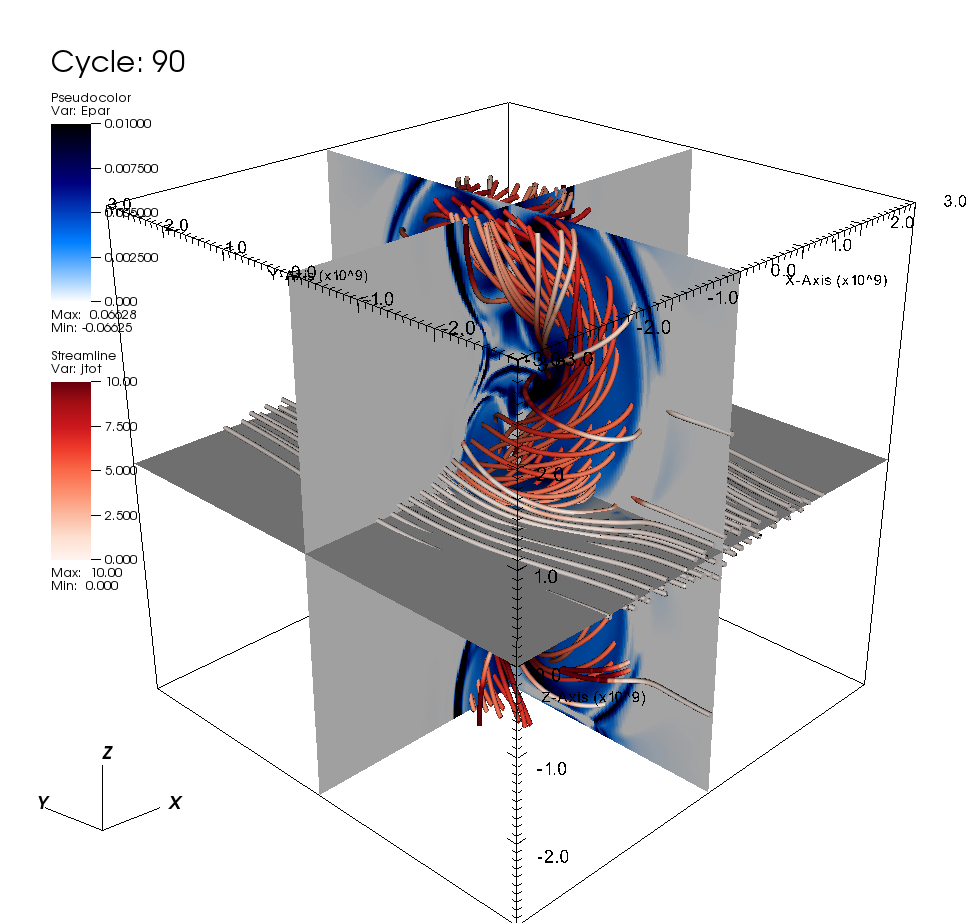}
\caption{3D view of the parallel, resistive electric field $E_{||}$ with slices cut through the three axes, initially at $t=0 t_S$ and in the nonlinear regime at $t=9t_S$ respectively. The colour scale is saturated at $0.001$. Selected magnetic field lines in the current channel area are shown, coloured by their total current density value, saturated at $j=10$. The box size is $6L\times 6L \times 6L$ with $L = 10^9 cm$.}
\label{fig:Epar3D}
\end{figure*}

In 3D the perturbation in the velocity field consists of a $z$-component and dependency. This introduces variations in the $z$-direction which is invariant in 2.5D setups. The repelling and rotating current channels develop an additional kink instability that causes reconnection. Strong and thin current sheets develop at the boundaries and in between the two repelling islands (see Fig.\ref{fig:Jtot2Dand3D} for the total current density magnitude in 2.5D and in 3D at $t = 9 t_S$). In \cite{Ripperda} it is shown that reconnection in this setup is indicated by a non-zero resistive electric field parallel to the magnetic field. This resistive electric field is plotted in Fig.~\ref{fig:Epar3D}, with selected reconnecting magnetic field lines, initially at $t = 0 t_S$ and far into the nonlinear regime at $t = 9t_S$. Particles mainly accelerate parallel to the resistive electric field and hence parallel to the magnetic field (\citealt{Ripperda}). In this section we investigate the effect of the kink instability on particle acceleration as well as the influence of the initial particle velocity distribution, the length of the flux ropes and resistivity for both electrons and protons. The effect of the initial spatial distribution and total number of particles on particle statistics are reported in Appendix \ref{sect:appnumber} -- \ref{sect:appspatial}. The effect of gyration is monitored in specific cases in Appendix \ref{sect:appgyration}.

\subsection{Effect of the kink instability on particle distributions}
\begin{figure*}
  \centering
   \subfloat{\includegraphics[width=\columnwidth, clip=true]{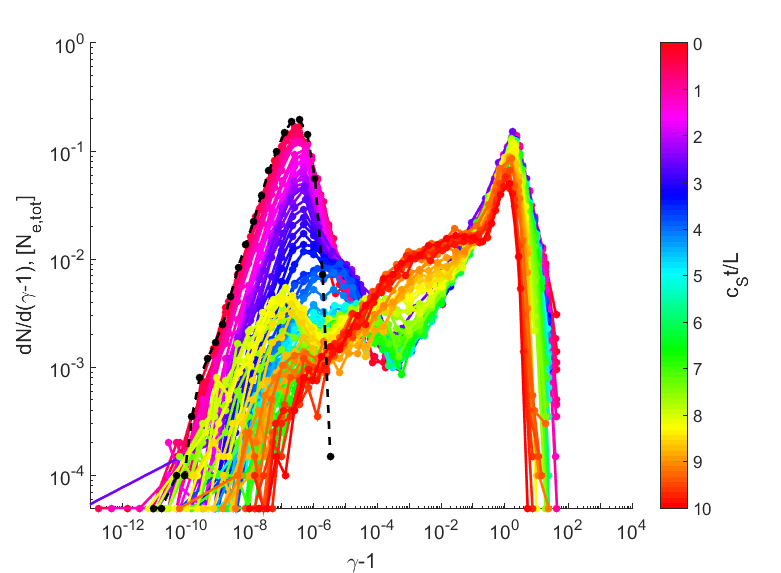}}
	   \subfloat{\includegraphics[width=\columnwidth, clip=true]{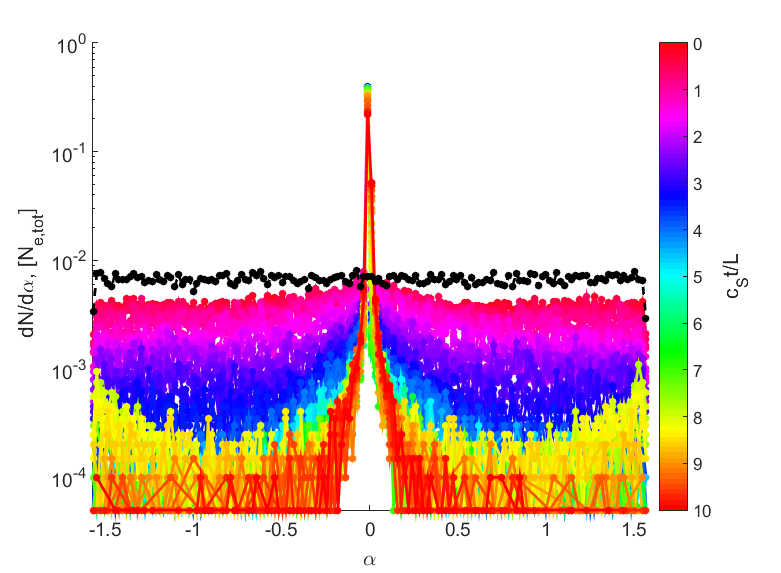}}
\caption{3D distributions for case A3De with 20.000 electrons with $\eta_p = 10^{-4}$ and $6L$ for the length of the current channels. Compared to 2.5D case A2De with equivalent settings in Fig.~\ref{fig:Ekin_distributions_20000electrons_2D} a medium energy tail, visible as a bump in between the two peaks, after $t \approx 8 t_S$. The particles are also redistributed in the thermal distribution at late times.}
\label{fig:Ekin_distributions_20000electrons_3D}
\end{figure*}

The effect of the kink instability is best visible for electrons. We show the kinetic energy spectra and the pitch angle spectra obtained from guiding centre simulations, with thermal velocity $v_{th,e} = v_{th,p} = \sqrt{2 k_B T \rho_0/(m_{p} p_0)}$ with initially a fraction of $0.99$ of the particles in the current channel area and flux ropes with length $6L$ in Fig.~\ref{fig:Ekin_distributions_20000electrons_3D} for 20.000 electrons (case A3De). The main observable difference due to the kink instability is the development of a medium energy tail in the electron energy distribution with $10^{-5} \leq \gamma -1 \leq 1$ starting at $t \approx 8 t_S$ in the nonlinear regime. A second noticeable effect is the redistribution of electrons in the thermal distribution $\gamma-1 \leq 10^{-5}$ after $t \approx 8 t_S$. This is attributed to the curvature of the magnetic field due to the kink, expelling particles from the current channels. These particles lose their energy in the ambient medium and either remain there and eventually leave the domain through the open boundaries, or they are caught again in either of the two current channels and re-accelerate. The electron energy distributions develop a high energy peak at $\gamma-1 \approx 60$ ($\sim 31$ MeV) and the high energy tail has an inverted power law index $p < 0$. The difference in maximum energy compared to 2.5D results can be explained by the peak current reached in the MHD evolution. The current sheet is narrower in 2.5D due to a higher resolution (\citealt{Ripperda}), resulting in a larger peak current that accelerates particles to higher energy.

\begin{figure*}
  \centering
	\subfloat{\includegraphics[width=\columnwidth, clip=true]{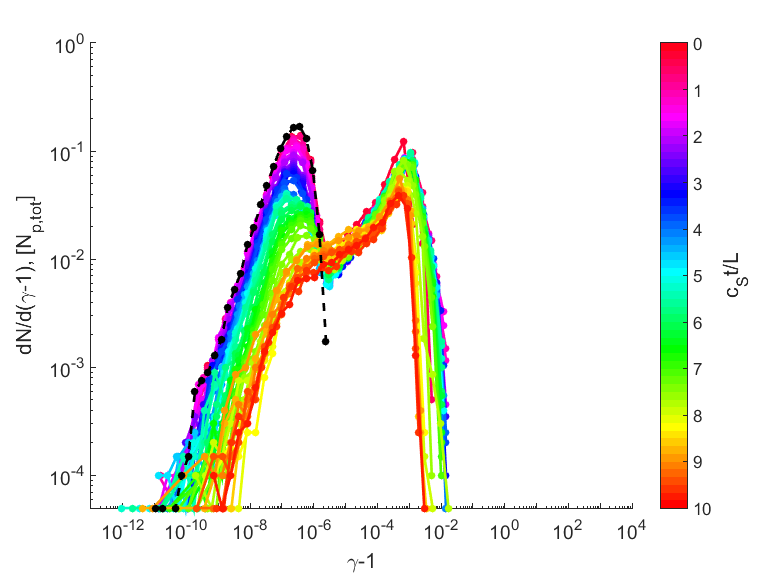}}
		\subfloat{\includegraphics[width=\columnwidth, clip=true]{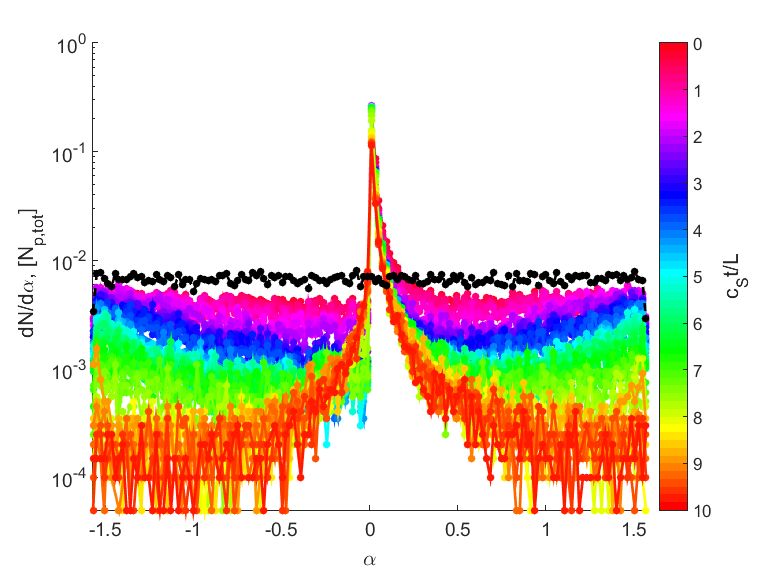}}
\caption{3D distributions for case B3Dp with 20.000 protons with $\eta_p = 10^{-4}$ and $6L$ for the length of the current channels. Compared to equivalent case A3De for electrons in Fig.\ref{fig:Ekin_distributions_20000electrons_3D} protons develop a less high maximum Lorentz factor and no medium energy tail forms. The asymmetry due to the acceleration of electrons and protons in opposite directions is more visible for the heavier protons in the pitch angle distribution on the right-hand panel.}
\label{fig:Ekin_distributions_20000protons_3D}
\end{figure*}

In Fig.~\ref{fig:Ekin_distributions_20000protons_3D} we show kinetic energy (left-hand panel) and pitch angle distributions (right-hand panel) for case B3Dp, with 20.000 protons with thermal speed $v_{th,p} = \sqrt{2 k_B T \rho_0/(m_{p} p_0)}$ in the same setup as electrons in case A3De.
The proton distributions are very similar to the 2.5D results. The maximum kinetic energy in the high energy peak for protons is $\gamma - 1 \lesssim 2 \cdot 10^{-2}$ or $\mathcal{E} \lesssim 957$ MeV and the high energy tail has an inverted power law index $p < 0$, similar to 2.5D case B2Dp. The pitch angle distributions are strongly peaked around $\alpha = 0$. However, the asymmetry is less pronounced in 3D. This is attributed to randomization of pitch angles due to 3D effects in the nonlinear regime after $t \approx 8 t_S$. The particles form two distinct populations. Protons and electrons moving inside the current channels develop a very high parallel velocity and a pitch angle very close to zero (positive for protons and negative for electrons). This is visible in the pitch angle distributions through the asymmetry around $\alpha = 0$ at early times (before the linear growth phase of the tilt-kink instability at $t \approx 5 t_S$). This population of particles is also observed in the kinetic energy distributions at $\gamma_{max} - 1 \approx 2 \cdot 10^{-2}$ for protons and $\gamma_{max} - 1 \approx 6 \cdot 10^{1}$ for electrons. During the nonlinear phase of the tilt-kink evolution ($t \gtrsim 5 t_S$) particles are expelled from the channels due to the kink and reach less high energies ($\gamma_{max} - 1 \approx 2 \cdot 10^{-3}$ for protons and $\gamma_{max} - 1 \approx 2 \cdot 10^{1}$ for electrons). In the pitch angle distributions this is observed through the more symmetric and less high peak around $\alpha = 0$. This is in accordance with \cite{Gordovskyy4} concluding that the width of the peak in the pitch angle distribution depends on the (maximum) particle energy. For higher maximum energy the particle species show a narrower peak. This is explained by taking into account that particles accelerate mostly along the magnetic field and drift velocities are negligible before reconnection occurs. Proton pitch angle distributions are wider than electron pitch angle distributions because electrons reach a larger parallel velocity and therefore the asymmetry at $\alpha = 0$ is more pronounced for protons.

\begin{figure*}
  \centering
		    \subfloat{\includegraphics[width=\columnwidth, clip=true, trim= 1cm 0cm 13cm 0cm]{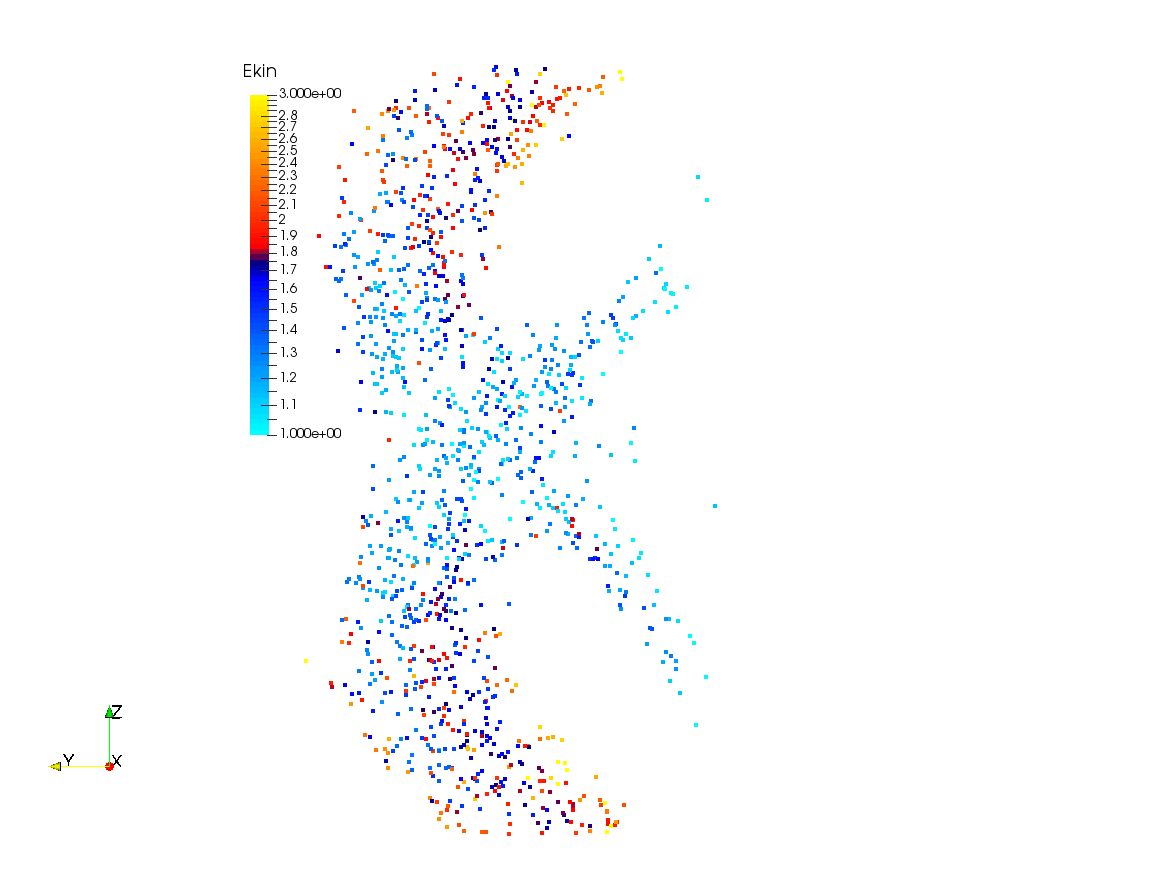}}
				\subfloat{\includegraphics[width=\columnwidth, clip=true, trim= 1cm 0cm 13cm 0cm]{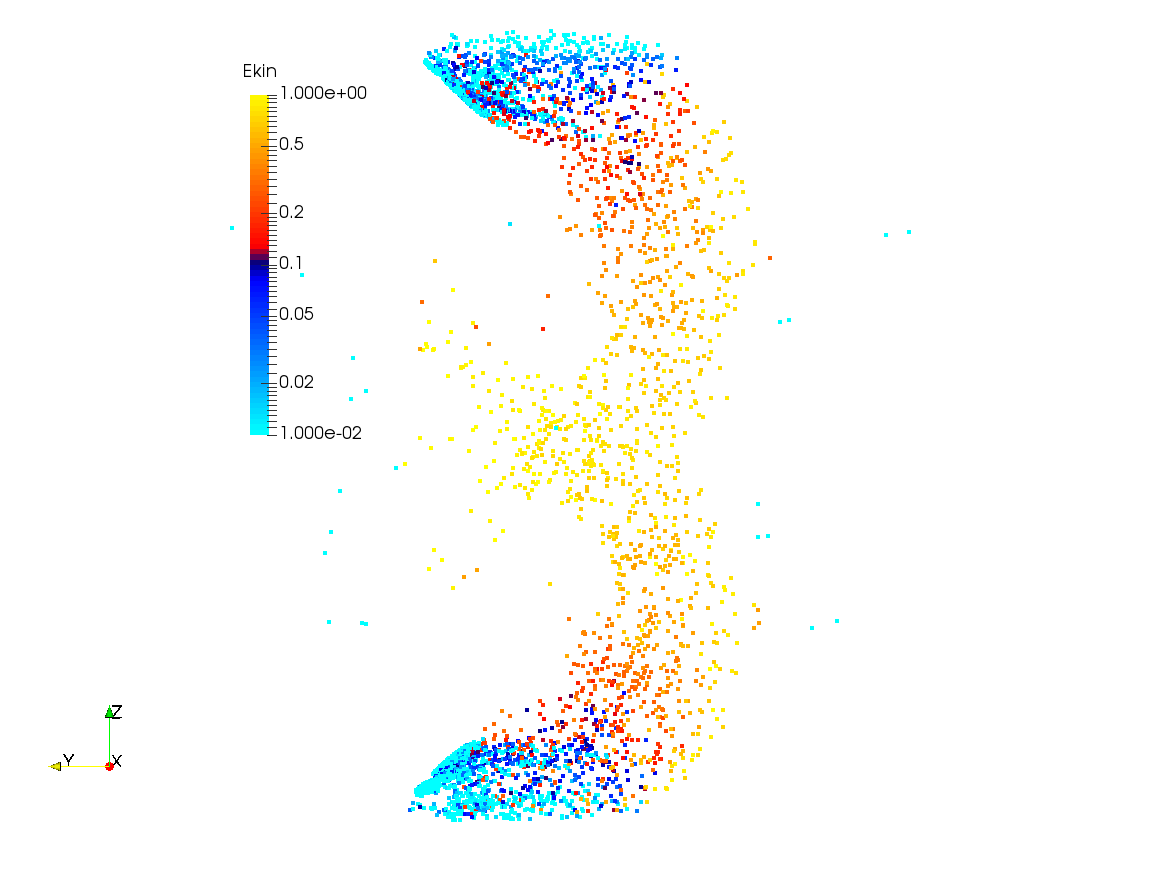}} 

    \subfloat{\includegraphics[width=\columnwidth, clip=true,  trim= 1cm 0cm 13cm 0cm]{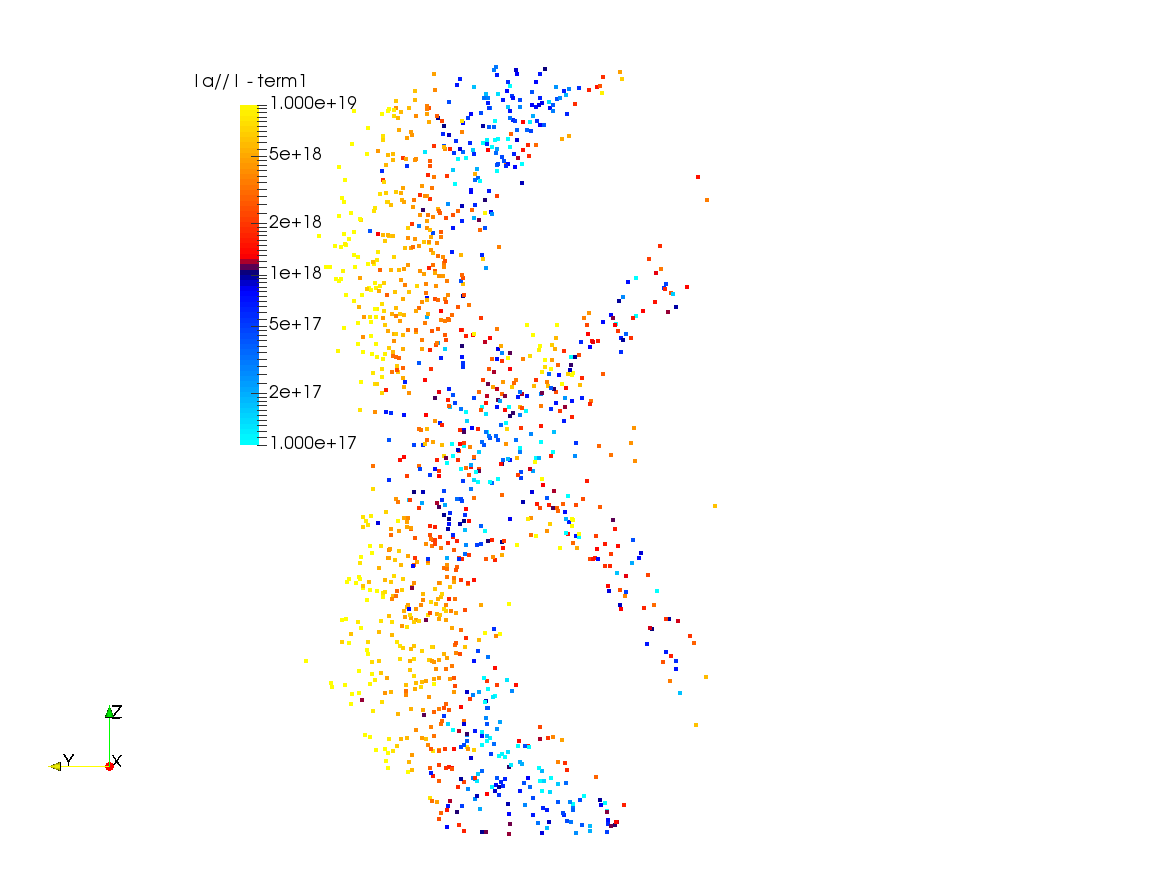}} 
		    \subfloat{\includegraphics[width=\columnwidth, clip=true,  trim= 1cm 0cm 13cm 0cm]{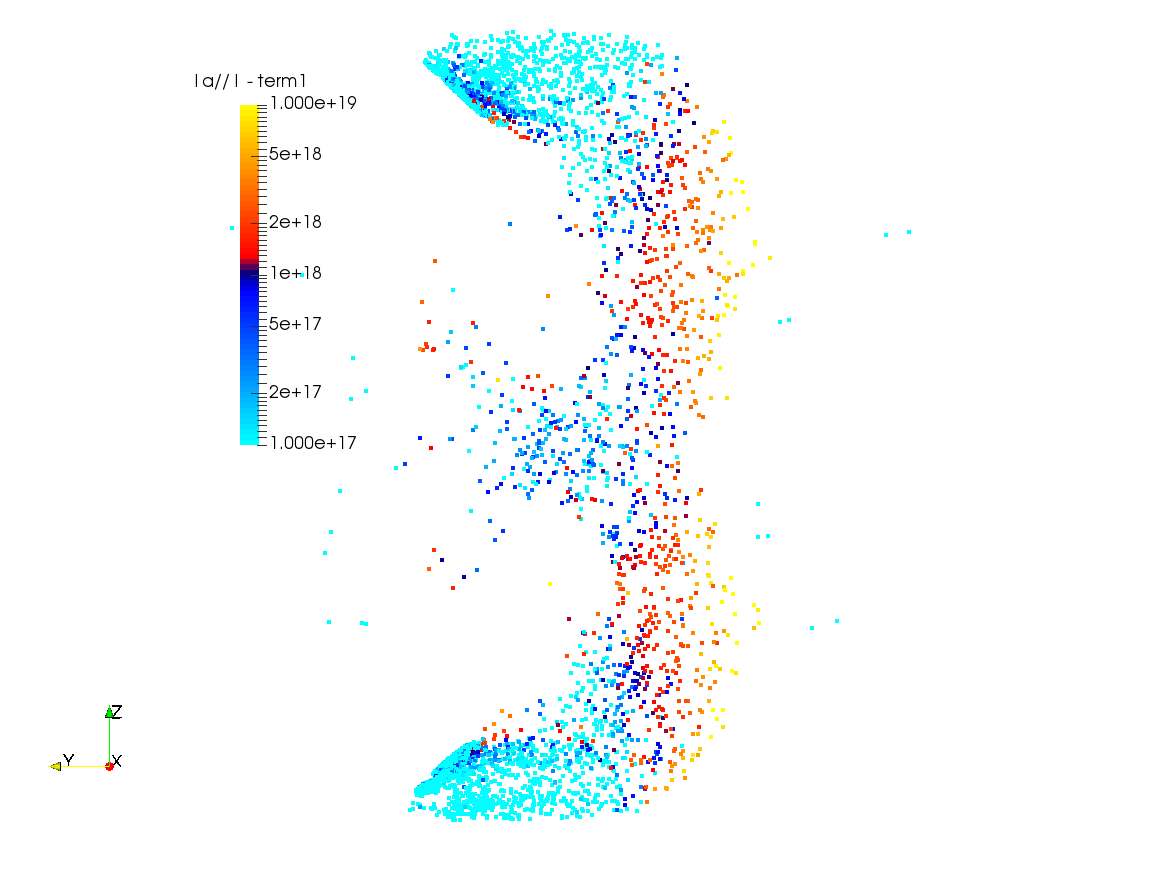}}
\caption{Spatial distribution of electrons at $t= 9t_S$ in case A3De, with thermal bath applied, particle resistivity $\eta_p = \eta_{MHD} = 10^{-4}$. In the panels on the left-hand-side high energy electrons, with $\gamma - 1 \geq 1$ are shown. In the top panel, electrons are coloured by their kinetic energy $E_{kin}/(m_0 c^2) = \gamma - 1$. In the bottom left-hand panel the same high energy electrons are coloured by the dominant contribution of the acceleration mechanism, the first term in equation (\ref{eq:gcastatic2}), the magnetic curvature magnitude $|\gamma\mathbf{u_E}\cdot \left(v_{\|}\left(\mathbf{\hat{b}}\cdot\nabla\right)\mathbf{\hat{b}}\right)|$. In the right-hand panels the low energy electrons, with $\gamma - 1 < 1$ are shown, coloured by their kinetic energy in the top panel and by the magnitude of the magnetic curvature acceleration in the bottom panel. The particles are projected in the $y,z$-plane with the line of sight along the $x$-direction. The particles follow magnetic field lines shown in \ref{fig:Epar3D} and at late times they reside mostly in the current channels, thus indicating the structure of the two current filaments.}
\label{fig:momentum_terms_spatial_eta4}
\end{figure*}

From the peaked pitch angle distributions in the right-hand panels of Figures \ref{fig:Ekin_distributions_20000electrons_3D} and \ref{fig:Ekin_distributions_20000protons_3D} we conclude that parallel acceleration is dominant. However, which effect causes this is not clear by just comparing the particle drifts from equation (\ref{eq:gcastatic1}). To analyze which acceleration mechanism causes this peak, the electrons are split in high energy particles, with $\gamma -1 \geq 1$ ($\mathcal{E} \gtrsim 0.5$ MeV), and low energy particles, with $\gamma - 1 < 1$. There are four contributions to the parallel acceleration in the momentum equation (\ref{eq:gcastatic2}); The first two $m_0\gamma\mathbf{u_E}\cdot \left(v_{\|}\left(\mathbf{\hat{b}}\cdot\nabla\right)\mathbf{\hat{b}}\right)$ and $m_0\gamma\mathbf{u_E}\cdot \left(\left(\mathbf{u_E}\cdot\nabla\right)\mathbf{\hat{b}}\right)$ due to the change of direction of the magnetic field (curvature and polarization effects respectively). The third $q E_{\|}$ due to resistive electric field and the last one $-\mu_r \mathbf{\hat{b}}\cdot\nabla\left[B\left(1-E^{2}_{\perp}/B^2\right)^{1/2}\right]/\gamma$ is the mirror deceleration effect. In Fig.~\ref{fig:momentum_terms_spatial_eta4} the spatial distribution of high energy electrons (left-hand panel) and the low energy electrons (right-hand panel) is shown, coloured by $(\gamma - 1)$ representing the particles kinetic energy (top panels) and by the curvature term $\gamma\mathbf{u_E}\cdot \left(v_{\|}\left(\mathbf{\hat{b}}\cdot\nabla\right)\mathbf{\hat{b}}\right)$ in the guiding centre momentum equation (\ref{eq:gcastatic2}) (bottom panel). This curvature term is found to be dominant at all time due to the initially curved magnetic field and the kink instability adding further curvature (see also further on in Section \ref{sect:resistivity} in Fig.\ref{fig:momentum_terms_t6}). In the left-hand channel (seen from the top), the particles move upwards, because of the direction of the magnetic field and hence the current density; In the right-hand channel, they move downwards. The particles are assigned a thermal speed every time they cross a $z$-boundary and therefore all fast particles are on the left of Fig.~\ref{fig:momentum_terms_spatial_eta4} and all the slow particles are on the right, with the thermal particles at the foot points. The magnetic curvature is equally divided between left and right channels. However the fast particles in the left-hand panel of Fig~\ref{fig:momentum_terms_spatial_eta4} traveled for a longer time in the current channels and obtained a larger acceleration and energy, mainly from the curvature. The curvature is stronger on the outside of the channels than on the inside and the fastest particles (indicated in red in the left-hand top panel) are therefore located at the foot points of the channels, on the outside, where also the current density is largest (\citealt{Ripperda}).

\subsection{Individual particle dynamics}
\label{sect:particles}
\begin{figure}
\centering
\subfloat{\includegraphics[width=\columnwidth, clip=true, trim= 1cm 0cm 13cm 0cm]{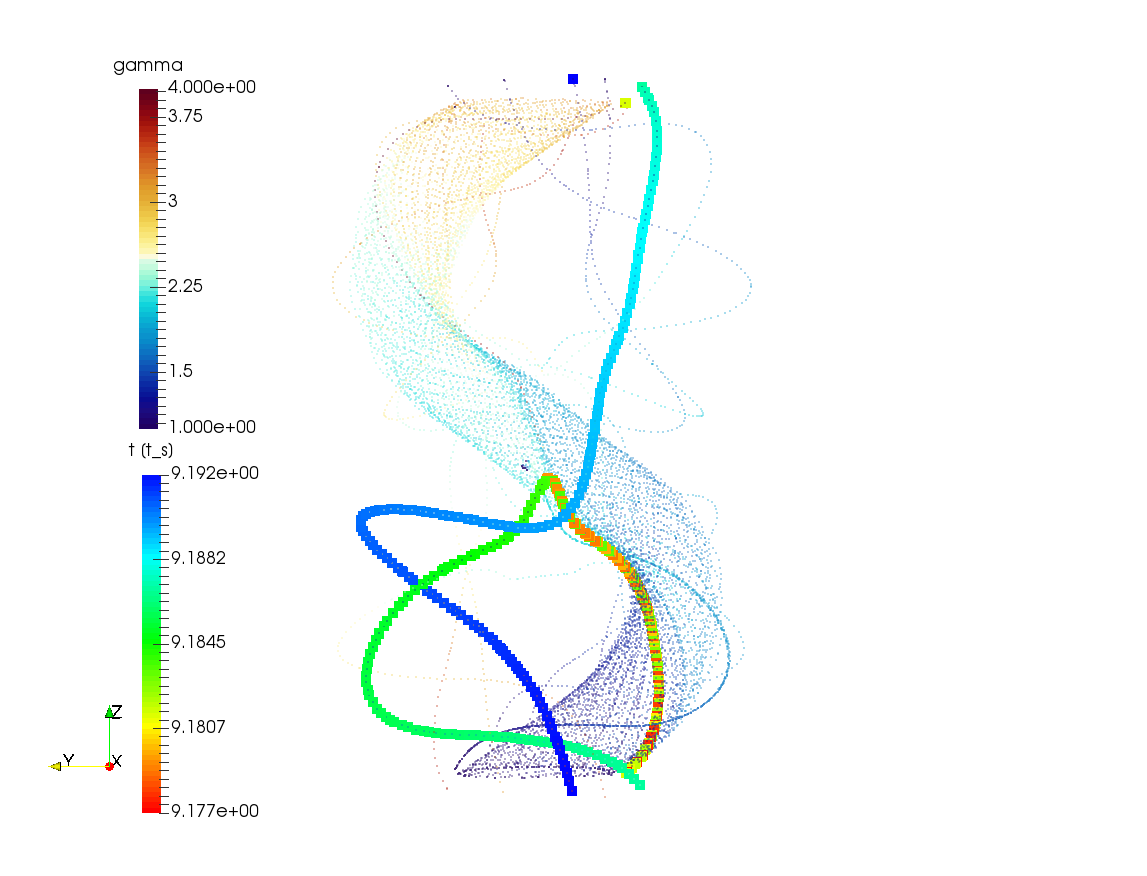}}
\caption{Side view in the $y,z$-plane of the trajectory of an electron that is expelled from the kinking current channel in case A3De in the time interval $[8.80t_S, 9.25 t_S]$. The trajectory is coloured by the particles Lorentz factor. Between $9.177t_S$ and $9.192 t_S$, when the particle is expelled from the flux rope the trajectory is marked with thicker squares and coloured by the time $t$ counted in $t_S$ indicated by the second legend. Thermal bath boundary conditions are applied at the top and bottom boundaries, where the fast particle is destroyed and a thermal particle is injected at the opposite boundary from a Maxwellian with thermal speed $v_{th,e} = \sqrt{2 k_B T \rho_0/(m_{p} p_0)}$ and $\gamma_{th,e} = 1$. It is clearly visible that the particle leaves the current channel at $t \approx 9.183 t_S$.}
\label{fig:trajectorie82823}
\end{figure}

To analyze how individual particles energize in the magnetic field of two interacting flux ropes we look at electron trajectories in run A3De. It is interesting to look at the energy evolution of a particle that is initially traveling in the current channel but is then expelled from the current channel. In Fig.~\ref{fig:trajectorie82823} we show the trajectory of such a particle from a side-view, coloured by its Lorentz factor. The electron cycles several times through the current channels until $t = 9.177 t_S$. From then onwards the trajectory is portrayed by thick squares coloured by the time t counted in $t_S$, until $t = 9.192 t_S$. 
\begin{figure*}
  \centering
		    \subfloat{\includegraphics[width=\columnwidth, clip=true]{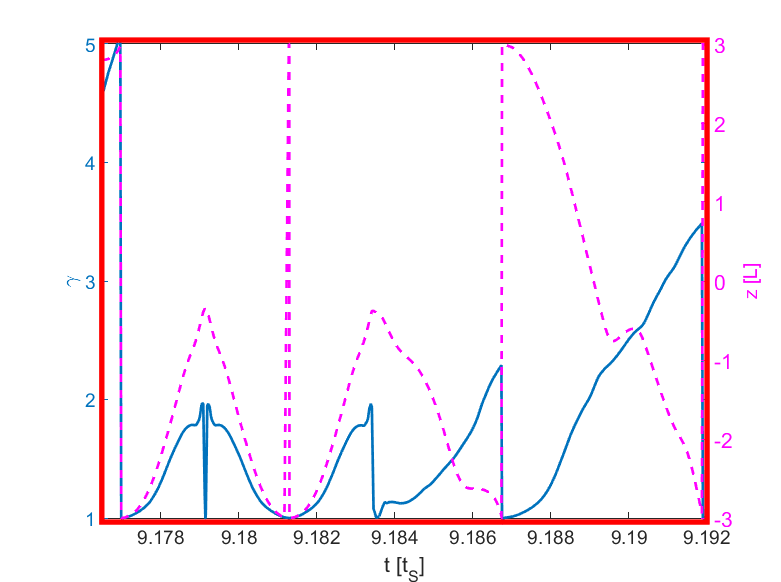}}
						    \subfloat{\includegraphics[width=\columnwidth, clip=true]{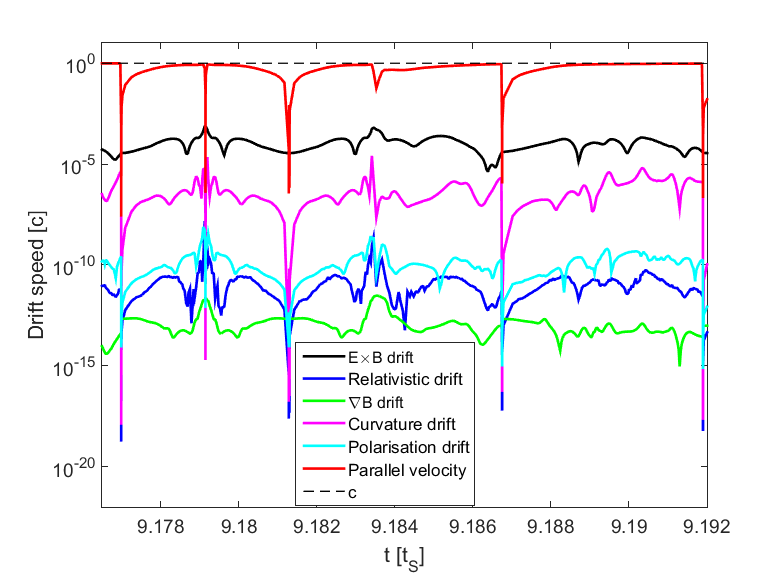}}
\caption{Temporal evolution of the Lorentz factor of the particle in Fig.~\ref{fig:trajectorie82823} zoomed into the time interval $t = 9.177t_S$ -- $t = 9.192 t_S$, (the period coloured by time $t$ in Fig.~\ref{fig:trajectorie82823}). To make a link with the complicated particles trajectory in Fig.~\ref{fig:trajectorie82823} the $z$-position of the particle is plotted with a dashed magenta line on the right-hand axis of the left-hand panel. At $t \approx 9.179 t_S$ the particle is mirrored and starts decelerating downwards in the channel. The particle is expelled from the current channel at $t \approx 9.183 t_S$. At $t \approx 9.181 t_S$ and $t \approx 9.187 t_S$ the particle reaches a boundary and is injected at the opposite boundary with a thermal velocity by the thermal bath. In the right-hand panel the temporal evolution of the magnitude of all drift terms on the right-hand-side of equation (\ref{eq:gcastatic1}) are shown, normalized to the speed of light (indicated by the black dashed line), during the same period.}
\label{fig:gamma82823}
\end{figure*}
This time interval corresponds to the red rectangle in the left-hand and the middle panel in Fig.~\ref{fig:gamma82823}, where the evolution of the Lorentz factor is shown. On the right-hand axis of the middle panel of Fig.~\ref{fig:gamma82823} the $z$-position of the particle is depicted with a magenta dashed line to indicate where the particle is on the trajectory in Fig.~\ref{fig:trajectorie82823}. The particle is first injected with a thermal speed at the bottom boundary at $z = -3$. The particle accelerates and travels upwards until it is mirrored by the magnetic field at $t \approx 9.179 t_S$. Then it starts traveling downwards towards $z = -3$ and decelerates. At $t \approx 9.181$ the particle reaches $z = -3$, it is injected at $z = 3$ with a thermal speed. It is then mirrored again towards $z = 3$. At $t \approx 9.181$ it leaves at $z = 3$ and is injected with a thermal speed at $z = -3$. It accelerates until it reaches a Lorentz factor of $\gamma \approx 2$ along the same trajectory upwards through the current channel. At $t = 9.1835$ it is expelled from the kinking current channel. Outside the current channel the particle moves downwards, following the magnetic field, until it reaches $z = -3$ at $t \approx 9.187 t_S$. It is injected again with a thermal speed at the top boundary at $z = 3$ and accelerates by following a field line outside the current channel. The particle Lorentz factor increases with time spent in the flux tube and the $z$-position increases (or decreases depending on the orientation) linearly in the current channels. In the right-hand panel of Fig.~\ref{fig:gamma82823} the evolution of the magnitude of the drift terms from the right-hand-side of equation (\ref{eq:gcastatic1}) is shown. The parallel velocity is dominant and quickly approaches the speed of light (black dashed line) when the particle accelerates. 

\subsection{Effect of resistivity on particle distributions}
\label{sect:resistivity}
\begin{figure*}
  \centering
   \subfloat{\includegraphics[width=\columnwidth, clip=true]{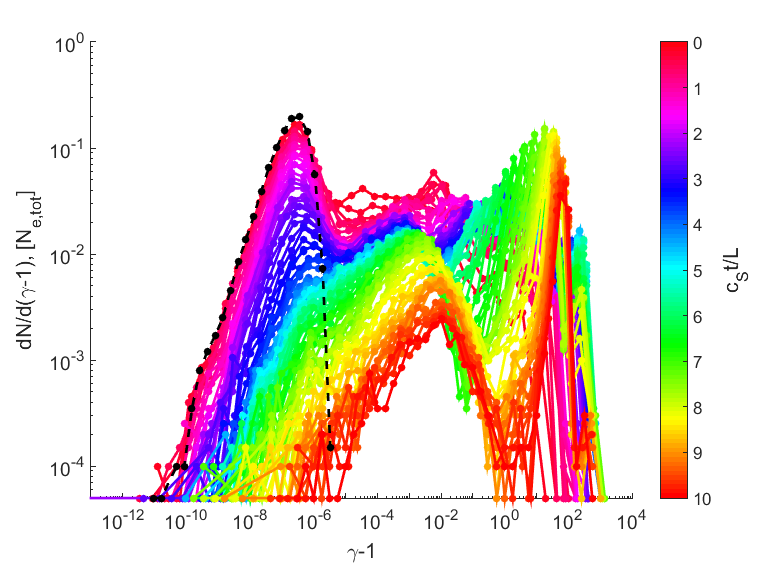}}
		    \subfloat{\includegraphics[width=\columnwidth, clip=true]{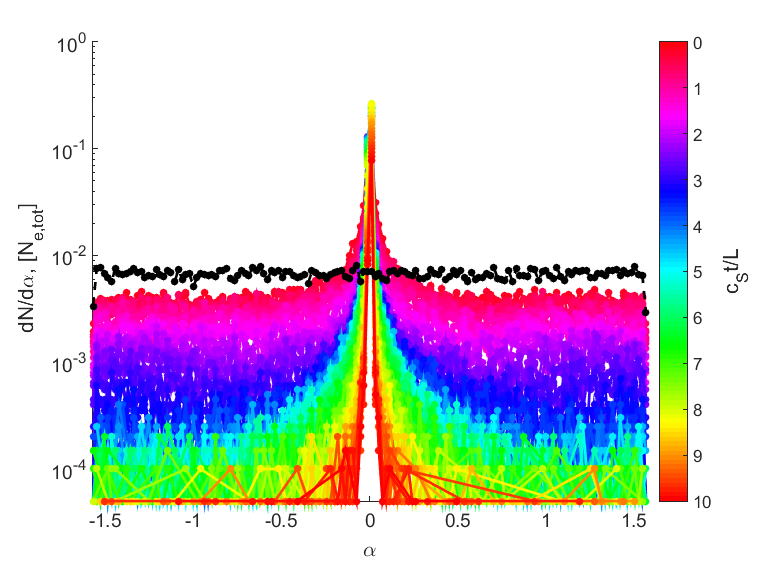}}
\caption{Distributions for case C3De with 20.000 electrons in 3D MHD with lowered resistivity $\eta_p = 10^{-9}$ and infinite length of the current channels (i.e. no thermal bath boundary condition applied). Compare to case A3De with higher resistivity and thermal bath in Fig.~\ref{fig:Ekin_distributions_20000electrons_3D}}.
\label{fig:Ekin_distributions_20000electrons_3D_smalleta10-9}
\end{figure*}

A resistivity of $\eta_p = 10^{-4}$ results in a magnetic Reynolds number of $R_m =2 \cdot 10^{5}$, which is three order of magnitude lower than in the solar corona. To moderate the acceleration parallel to the magnetic field, and therewith the peaked pitch angle spectra and the hard, inverted energy spectra, a resistivity model is proposed that lowers the resistivity to realistic solar corona values. To restrict the contribution of a resistive electric field the resistivity in the GCA equations is set to $\eta_p=10^{-9}$ resulting in a Magnetic Reynolds number $R_m =2 \cdot 10^{10}$ in the solar corona regime. The energy spectra for case C3De with $\eta_p = 10^{-9}$ and periodic (infinite) flux ropes are shown in the left-hand panel of Fig.~\ref{fig:Ekin_distributions_20000electrons_3D_smalleta10-9} and the pitch angle spectra in the right-hand panel. A high energy peak develops at $t \approx 5 t_S$ with maximum Lorentz factor $\gamma -1\lesssim 10^3$, resulting in $\mathcal{E}_{max} \approx 500$ MeV for electrons and a medium energy part with an inverted slope. The pitch angle is still strongly peaked at $\alpha=0$ and the energy spectra still show an inverted slope.

\begin{figure*}
  \centering
\subfloat{\includegraphics[width=\columnwidth, clip=true]{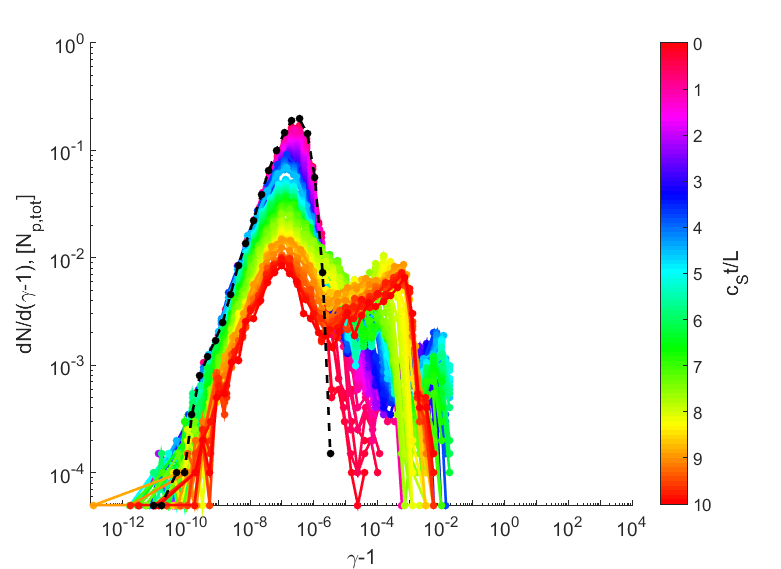}}   
 \subfloat{\includegraphics[width=\columnwidth, clip=true]{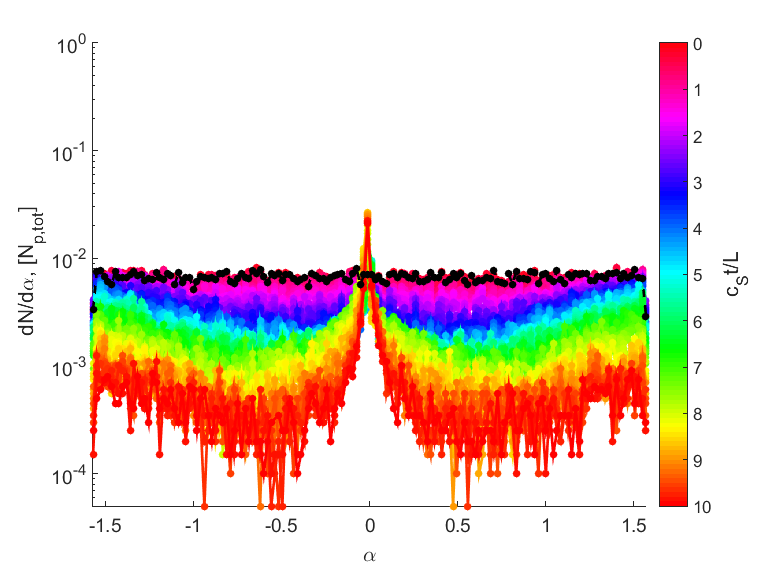}}
\caption{Distributions for case D3Dp with 20.000 protons uniformly distributed in 3D MHD with lowered resistivity $\eta_p = 10^{-9}$ and infinite length of the current channels (i.e. no thermal bath applied). Compare to proton case B3Dp with higher resistivity and thermal bath in Fig.~\ref{fig:Ekin_distributions_20000protons_3D}.}
\label{fig:Ekin_distributions_20000protons_3D_smalleta10-9_uniform}
\end{figure*}

In case D3Dp 20.000 protons are evolved in similar settings as case C3De for electrons. We find a final kinetic energy distribution (see left-hand panel of Fig.~\ref{fig:Ekin_distributions_20000protons_3D_smalleta10-9_uniform}) with maximum Lorentz factor $\gamma -1 \leq 2 \cdot 10^{-2}$ ($\mathcal{E} \lesssim 957$ MeV) and an inverted power law slope. The pitch angle spectrum (see right-hand panel of Fig.~\ref{fig:Ekin_distributions_20000protons_3D_smalleta10-9_uniform}) is peaked around $\alpha = 0$ but the peak is less dominant than in case B3Dp, where $\eta_p = \eta_{MHD} = 10^{-4}$ but the length of the flux ropes is limited to $6L$. 

\begin{figure*}
  \centering
      \subfloat{\includegraphics[width=\columnwidth, clip=true,  trim= 1cm 0cm 13cm 0cm]{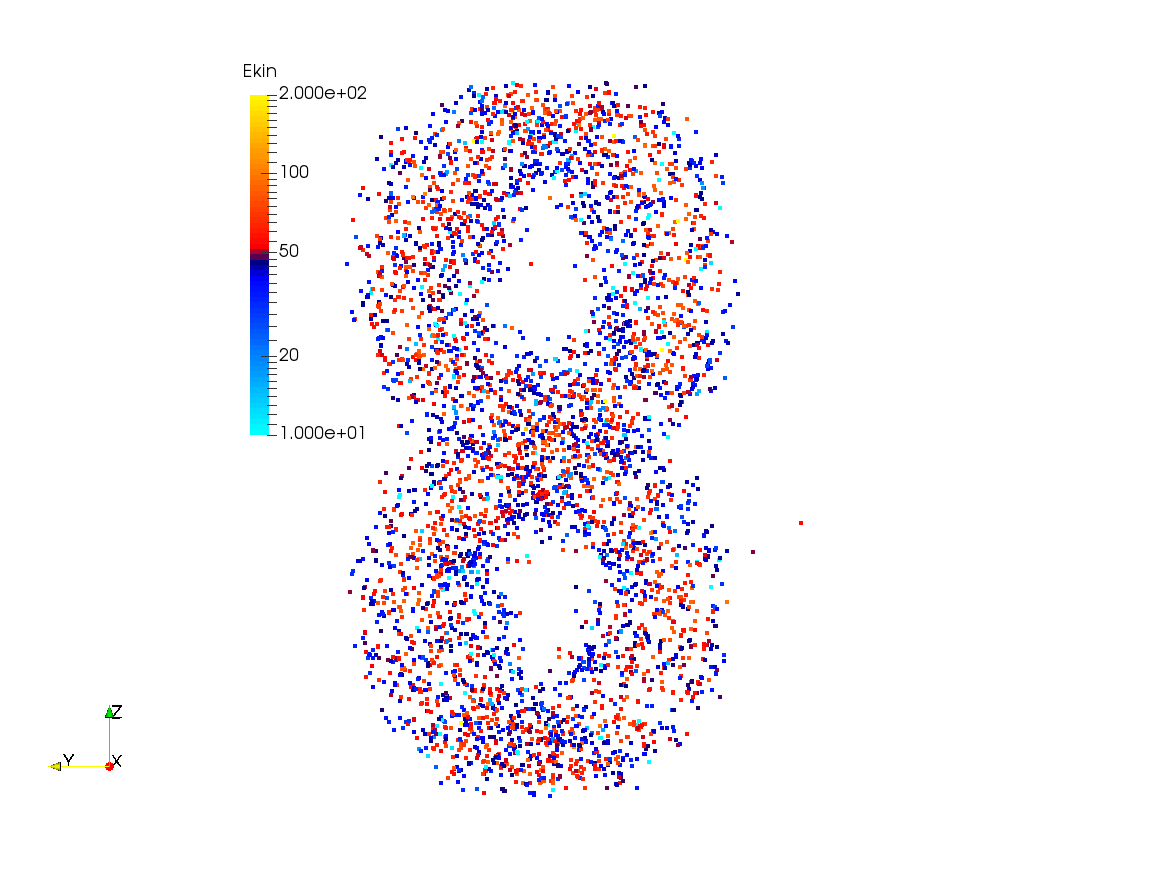}}
		\subfloat{\includegraphics[width=\columnwidth, clip=true,  trim= 1cm 0cm 13cm 0cm]{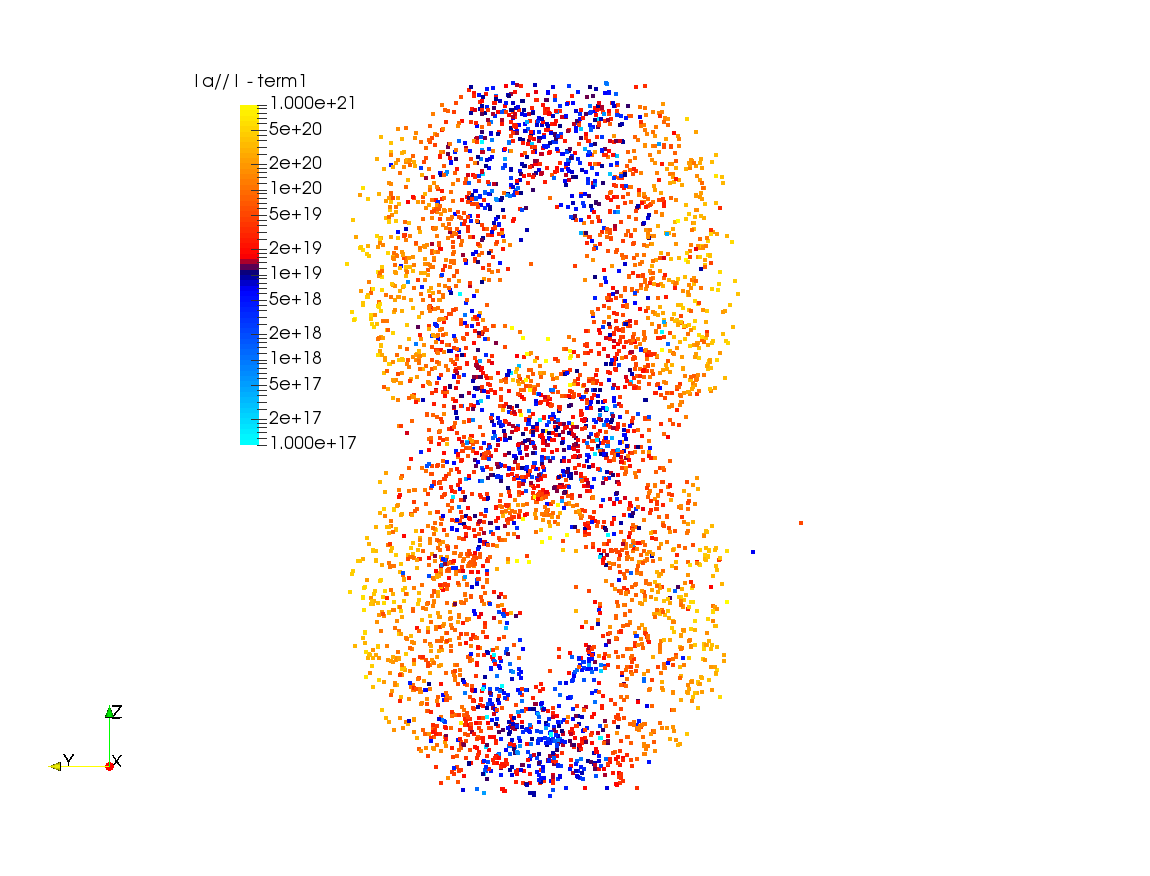}}
\caption{Spatial distribution of electrons at $t= 9t_S$ in case C3De, without thermal bath, particle resistivity $\eta_p = 10^{-5} \times \eta_{MHD} = 10^{-9}$. In the left-hand panel high energy electrons, with $\gamma - 1 \geq 1$ are coloured by their kinetic energy $E_{kin}/(m_0 c^2) = \gamma - 1$. In the right-hand panel the same high energy electrons are coloured by the dominant contribution of the acceleration mechanism, the first term in equation (\ref{eq:gcastatic2}), the magnetic curvature magnitude $|\gamma\mathbf{u_E}\cdot \left(v_{\|}\left(\mathbf{\hat{b}}\cdot\nabla\right)\mathbf{\hat{b}}\right)|$. The particles are projected in the $y,z$-plane with the line of sight along the $x$-direction. The particles follow magnetic field lines shown in \ref{fig:Epar3D} and at late times they reside mostly in the current channels, thus indicating the structure of the two current filaments.}
\label{fig:momentum_terms_spatial_eta9}
\end{figure*}

The high energy particles with $\gamma -1 \geq 1$ ($\mathcal{E} \gtrsim 0.5$ MeV) in case C3De are depicted in Fig.~\ref{fig:momentum_terms_spatial_eta9}, coloured by the kinetic energy $(\gamma-1)$ in the left-hand panel and by the magnitude of the curvature acceleration $m_0\gamma\mathbf{u_E}\cdot \left(v_{\|}\left(\mathbf{\hat{b}}\cdot\nabla\right)\mathbf{\hat{b}}\right)$ in the right-hand panel. Unlike for case A3De (compare to Fig.~\ref{fig:momentum_terms_spatial_eta4}), the fast particles are not just located at the foot points, but distributed over the whole area of the current channels, with the fastest particles in the middle (see the left-hand top panel). This area corresponds to the region with the strongest current density (and hence resistive electric field) and magnetic field curvature (see Fig.~\ref{fig:Epar3D} for the resistive electric field with magnetic field lines in this setup and \cite{Ripperda} for more detail on MHD results). The slow particles, with $\gamma - 1 < 1$, are residing outside the current channels. Particles mostly accelerate in the regions with strongest curvature, at the outside of the kinked channels. This can also be seen from the distribution of the acceleration terms in the bottom panel of Fig.~\ref{fig:momentum_terms_t6}. For case C3De the distribution of curvature acceleration at $t=9 t_S$ shows a tail consisting of the fast particles (in magenta with crosses as indicators). Compared to case A3De (Fig.~\ref{fig:momentum_terms_spatial_eta4}) there is a clearer separation between slow and fast particles and the contribution of curvature acceleration to the fast particles. In case A3De (Fig.~\ref{fig:momentum_terms_spatial_eta4}), all particles are accelerated by the magnetic curvature and the thermal bath prevents a tail to arise in the acceleration distribution.

The limited length of the flux ropes and the realistic particle resistivity separately counteract the strong parallel acceleration and the hard spectra found by \cite{Ripperda}; These two solutions are compared by quantifying the contributions of the four separate terms in the momentum equation (\ref{eq:gcastatic2}). We can determine whether the peaked distributions have a physical cause or are due to a too high MHD resistivity set for computational purpose. In Fig.~\ref{fig:momentum_terms_t6} the distribution of the four terms contributing to the particle momentum in the momentum equation (\ref{eq:gcastatic2}) are shown for case C3De and A3De at $t=6$, before the current channels start kinking and at $t=9$, in the nonlinear regime, respectively. We see that at $t=6$ the acceleration mechanisms show a similar distribution, compared between runs with thermal bath and $\eta_p = 10^{-4}$ (case A3De) and the runs with periodic boundary conditions and $\eta_{p} = 10^{-9}$ (case C3De), except the term due to the resistive, parallel electric field. For this term, $q E_{\|}/m_0 = q \eta_p \mathbf{J} \cdot \mathbf{\hat{b}}/m_0$ we can see the direct effect of decreased resistivity. At $t=9$ however, in the bottom panel of Fig.~\ref{fig:momentum_terms_t6}, the distributions of the curvature acceleration contribution $\gamma\mathbf{u_E}\cdot \left(v_{\|}\left(\mathbf{\hat{b}}\cdot\nabla\right)\mathbf{\hat{b}}\right)$ and the acceleration due to polarization $\gamma\mathbf{u_E}\cdot \left(\left(\mathbf{u_E}\cdot\nabla\right)\mathbf{\hat{b}}\right)$, are shifted by two orders of magnitude for case C3De with $\eta_{p} = 10^{-9}$. For the case with thermal bath (A3De) they are not. The mirror effect $-\mu_r \mathbf{\hat{b}}\cdot\nabla\left[B\left(1-E^{2}_{\perp}/B^2\right)^{1/2}\right]/\gamma$ is negligible in both cases C3De and A3De. The resistive acceleration $qE_{\|}/m_0$ has not changed with respect to the spectrum before the channels started kinking at $t=6 t_S$. Conclusively, the resistive electric field is not the main cause for the hard spectra obtained. The periodic boundary conditions and therewith the indefinite acceleration in the $z$-direction, in the current channels seem to be dominant and this effect is counteracted by thermal bath boundary conditions. However, even with a thermal bath applied, an inverted slope is observed in the energy spectra. 
\begin{figure*}
  \centering
	\subfloat{\includegraphics[width=2.3\columnwidth, trim= 3cm 0cm 0cm 0cm, clip=true]{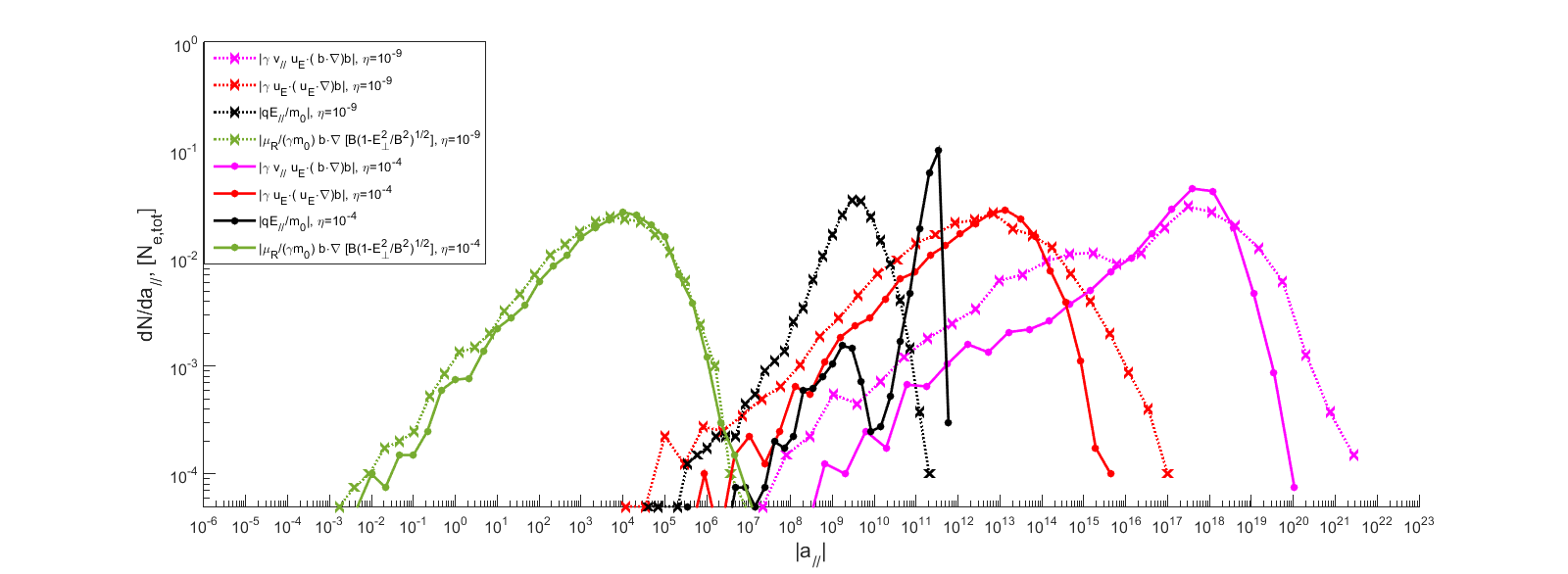}}
	
	\subfloat{\includegraphics[width=2.3\columnwidth, trim= 3cm 0cm 0cm 0cm, clip=true]{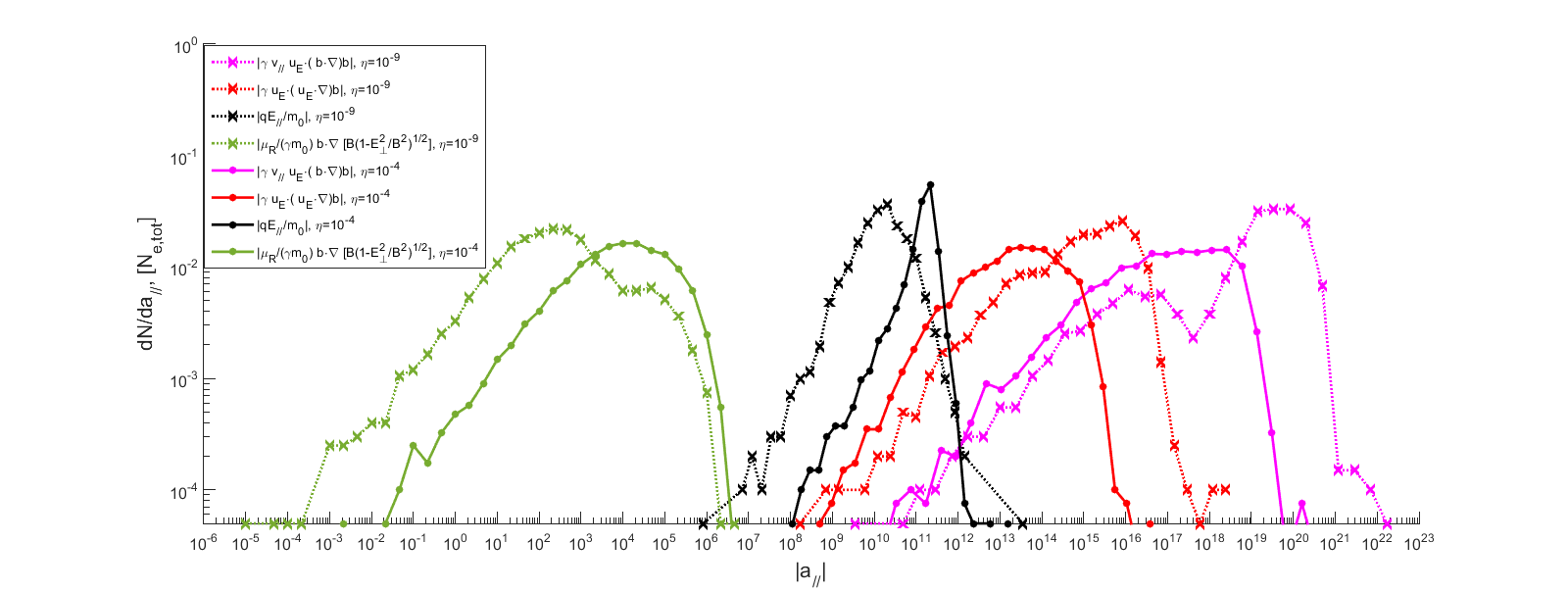}}
\caption{Comparison of the number distribution of the four acceleration terms in the right-hand-side of equation (\ref{eq:gcastatic2}) for 20.000 electrons in case C3De (dotted lines with crosses as indicators) and case A3De (solid lines with dots as indicators) before nonlinear phase at $t=6 t_S$ (top panel) and during nonlinear phase at $t=9 t_S$ (bottom panel). In case C3De periodic boundary conditions are applied (i.e. infinitely long flux ropes) and $\eta_p = 10^{-5} \times \eta_{MHD} = 10^{-9}$ and in case A3De the flux rope length is $6L$ and resistivity is set $\eta_p = \eta_{MHD} = 10^{-4}$. The magnetic mirror term (fourth term in equation \ref{eq:gcastatic2}) is indicated by a green line, the resistive electric field acceleration (third term in equation \ref{eq:gcastatic2}) by a black line and the two terms attributed to the change of direction of the magnetic field in red (polarization term, the second term in equation \ref{eq:gcastatic2}) and in magenta (magnetic curvature term, the first term in equation \ref{eq:gcastatic2}).}
\label{fig:momentum_terms_t6}
\end{figure*}

\subsection{Realistic conditions for solar flares}
\begin{figure*}
  \centering
    \subfloat{\includegraphics[width=\columnwidth, clip=true]{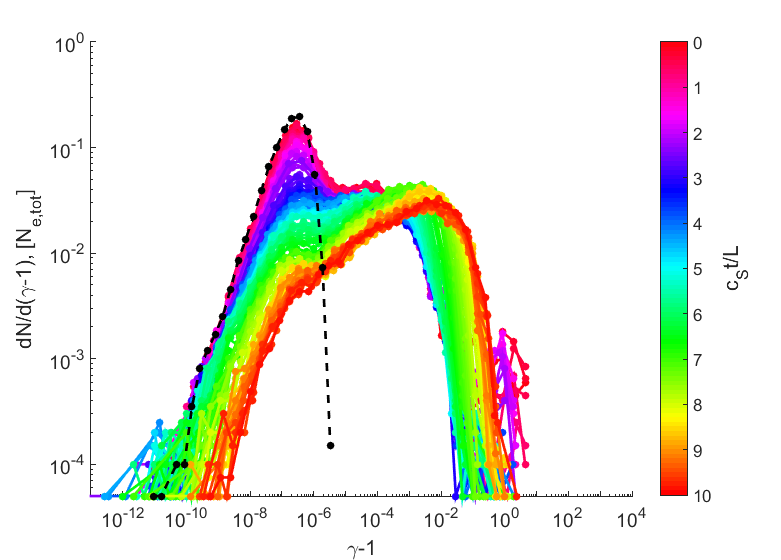}}
				    \subfloat{\includegraphics[width=\columnwidth, clip=true]{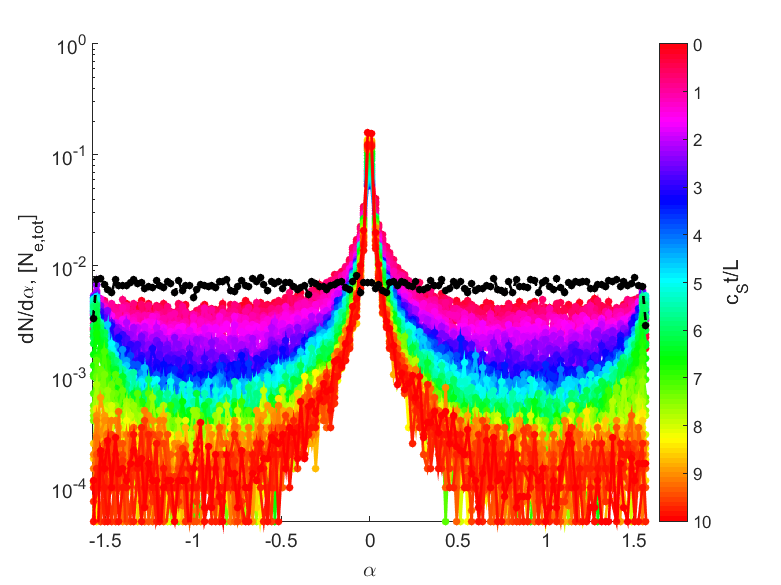}}
\caption{Distributions for case E3De with 20.000 electrons in 3D MHD with lowered resistivity $\eta_p = 10^{-9}$ and a limited length of $6L$ for the length of the current channels (i.e. thermal bath applied). Compare to cases A3De with higher resistivity (Fig.~\ref{fig:Ekin_distributions_20000electrons_3D}) and C3De with infinitely long current channels and lowered resistivity $\eta_p = 10^{-9}$ (Fig.~\ref{fig:Ekin_distributions_20000electrons_3D_smalleta10-9}).}
\label{fig:Ekin_distributions_20000electrons_3D_smalleta10-9_thermalisation}
\end{figure*}

In case E3De we combine a limited flux rope length and resistivity for realistic solar corona conditions. In a flux rope of length $6L$ and a resistivity of $\eta_p = 10^{-9}$ we evolve 20.000 electrons, with the aim to restrict a high energy peak and the accompanying inverted power law index for the high energy tail. The pitch angle distributions (right-hand panel of Fig.~\ref{fig:Ekin_distributions_20000electrons_3D_smalleta10-9_thermalisation}) are still dominated by $\alpha = 0$, but there are more particles with a nonzero pitch angle, compared to cases A3De and C3De. The high energy peak observed in case A3De and C3De has now disappeared in the kinetic energy distributions (left-hand panel of Fig.~\ref{fig:Ekin_distributions_20000electrons_3D_smalleta10-9_thermalisation}). The slope of the high-energy tail at times $t \gtrsim 7 t_S$ (coloured green to red), when the channels start kinking, is still inverted. However, at this time there are not many particles left in the thermal part of the distribution. The maximum kinetic energy is bounded by $\gamma - 1 \lesssim 5$ ($\mathcal{E} \lesssim 3$ MeV), due to the thermal bath and the length of the channels. How many particles remain thermal is affected by the initial velocity distribution. The particles maximum energy is bounded by the time the particles spend in the flux rope, and hence by the length of the flux rope. 

\subsection{Effect of initial velocity distribution on particle distributions}
\begin{figure*}
  \centering
    \subfloat{\includegraphics[width=\columnwidth, clip=true]{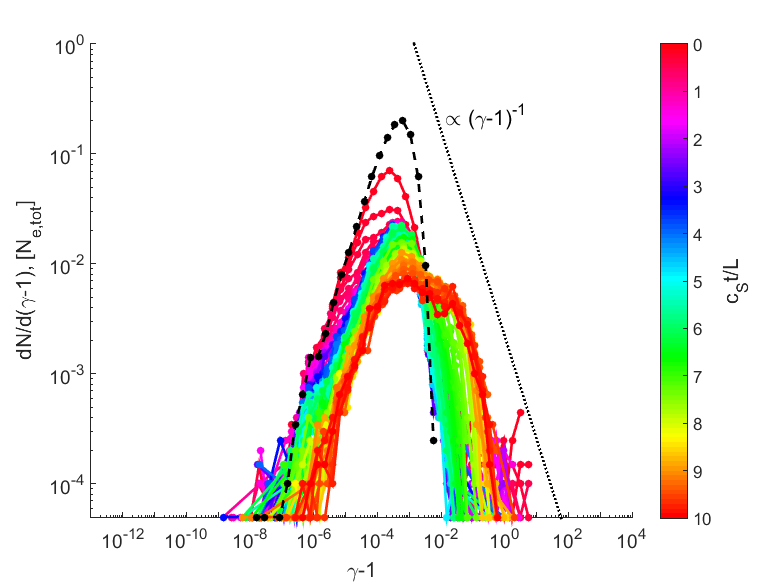}}
		    \subfloat{\includegraphics[width=\columnwidth, clip=true]{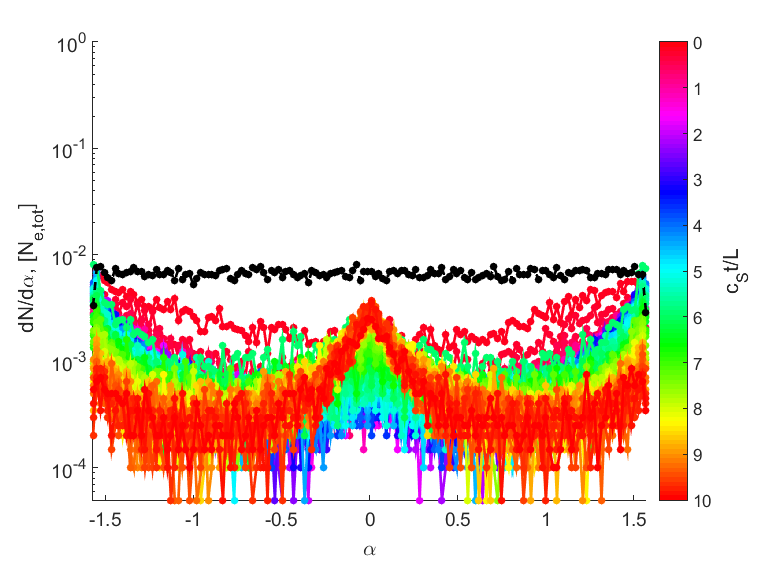}}
\caption{Distributions for 20000 electrons in case G3De with an initially uniform spatial distribution with thermal velocity $v_{th,e}=\sqrt{2 k_B T \rho_0/(m_{p} p_0)}\sqrt{m_{p}/m_{e}}$ and lowered resistivity $\eta_p = 10^{-9}$ and $6L$ for the length of the current channels (i.e. thermal bath applied). Compare to case E3De (Fig.~\ref{Ekin_distributions_20000electrons_3D_smalleta10-9_thermalisation}) with an initially smaller thermal velocity $v_{th,e} = v_{th,p}$ for electrons. As a guide for the eye a typical power law distribution of the form $dN/d(\gamma-1) \propto (\gamma-1)^{-1}$ is plotted (dotted black line), corresponding to equal energy content in each decade of $\gamma-1$.}
\label{fig:Ekin_distributions_20000electrons_3D_smalleta10-9_uniform_massratio}
\end{figure*}

Up to now it is assumed that all particles, electrons and protons, have the typical MHD velocity $v_{th,p}=\sqrt{2 k_B T \rho_0/(m_{p} p_0)}$ as thermal speed. Assuming energy equipartition between electrons and protons results in a larger thermal speed for electrons $v_{th,e}= \sqrt{m_p/m_e} \times v_{th,p}$. In case G3De we explore the effect of a larger initial electron velocity with flux rope length $6L$ and particle resistivity set as $\eta_p = 10^{-9}$. Electrons are initialized from a Maxwellian with thermal speed $v_{th,e} = \sqrt{m_p/m_e} \times v_{th,p}$ in accordance with energy equipartition and they are uniformly distributed over the spatial domain $ -3L \leq x \leq 3L$; $-3L \leq y \leq 3L$; $ -3L \leq z \leq 3L$. Few particles accelerate inside the current channels at early times $t < 6 t_S$ in the linear phase (see Fig.~\ref{fig:Ekin_distributions_20000electrons_3D_smalleta10-9_uniform_massratio}). The maximum electron energy is limited by the length of the current channels and the thermal bath to $\gamma - 1 \lesssim 7$ or $\mathcal{E} \lesssim 4.1$ MeV, which is in the range of observed electron energies coming from solar flares. New thermal particles are injected from the thermal bath for every destroyed high energy particle leaving a periodic boundary, maintaining a thermal distribution dominant in number of particles. The high energy tail that develops during the exponential growth phase of the tilt-kink instability after $t \geq 6 t_S$ has a power law distribution with spectral index $p > 1$ (indicated by the dotted black line in the left-hand panel of Fig.~\ref{fig:Ekin_distributions_20000electrons_3D_smalleta10-9_uniform_massratio}). The high energy particles are not dominant in number of particles, nor in energy content due to decreased particle resistivity and the thermal bath. The pitch angle distributions in the right-hand panel of Fig.~\ref{fig:Ekin_distributions_20000electrons_3D_smalleta10-9_uniform_massratio} are initially (nearly) uniform and the peak due to particles accelerated parallel to the magnetic field at $\alpha = 0$ develops only after $t \geq 6 t_S$. The peak at $\alpha = 0$ is less dominant than in all other electron cases.

\begin{figure*}
  \centering
\subfloat{\includegraphics[width=\columnwidth, clip=true]{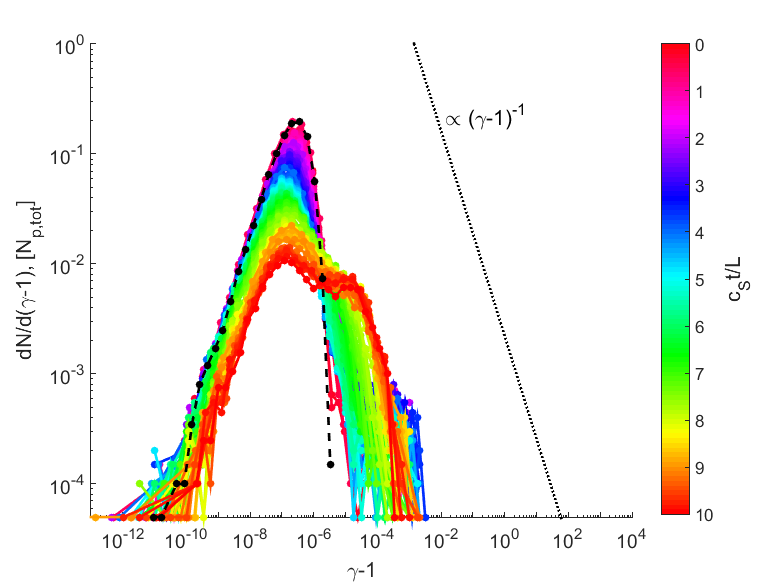}}   
 \subfloat{\includegraphics[width=\columnwidth, clip=true]{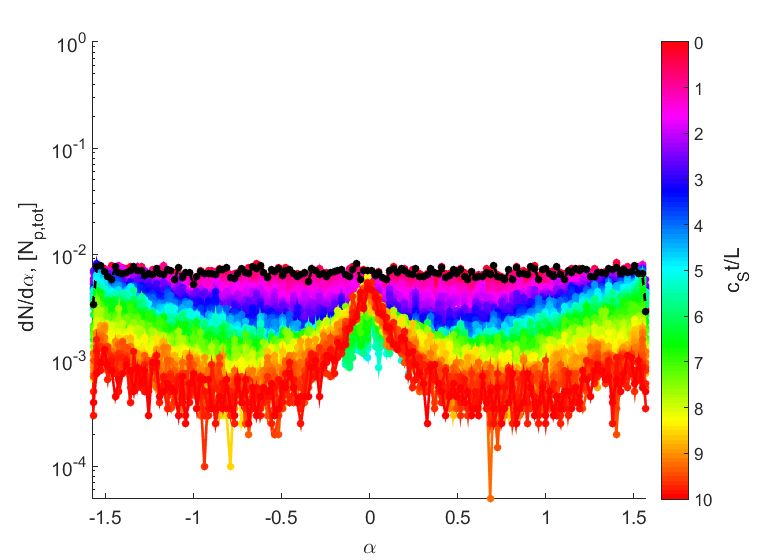}}
\caption{Distributions for 20000 protons in case F3Dp with an initially uniform spatial distribution and lowered resistivity $\eta_p = 10^{-9}$ and $6L$ for the length of the current channels. Compare to cases B3Dp (Fig.~\ref{fig:Ekin_distributions_20000protons_3D}) with infinitely long current channels and lowered resistivity $\eta_p = 10^{-9}$ and D3Dp (Fig.~\ref{fig:Ekin_distributions_20000electrons_3D_smalleta10-9}) with thermal bath applied and higher resistivity $\eta_p = 10^{-4}$. As a guide for the eye a typical power law distribution of the form $dN/d(\gamma-1) \propto (\gamma-1)^{-1}$ is plotted (dotted black line), corresponding to equal energy content in each decade of $\gamma-1$.}
\label{fig:Ekin_distributions_20000protons_3D_smalleta10-9_uniform_thermalbath}
\end{figure*}

In case F3Dp we explore the same configuration for protons, initiated uniformly from a Maxwellian with $v_{th,p}$. For protons assuming energy equipartition or a generic fluid velocity as thermal speed results in the same initial energy distribution. The results are similar to electron case G3De with the difference that the maximum energy is limited to $\gamma -1 \lesssim 3 \cdot 10^{-2}$ or $\mathcal{E} \lesssim 941$ MeV (see the left-hand panel of Fig.~\ref{fig:Ekin_distributions_20000protons_3D_smalleta10-9_uniform_thermalbath}) due to the mass difference and consequently the lower thermal speed. Few particles accelerate inside the current channels at early times $t < 6 t_S$ in the linear phase. In the nonlinear phase from $t > 6 t_S$ onwards, a high energy tail develops with a power law distribution with index $p > 1$ (indicated by the dotted black line in the left-hand panel of Fig.~\ref{fig:Ekin_distributions_20000protons_3D_smalleta10-9_uniform_thermalbath}). Because of the lower maximum energy reached, less protons leave the domain through the open $x,y$-boundaries, compared to electrons in case G3De and the thermal distribution remains dominant at all times. The pitch angle distribution (see the right-hand panel of Fig.~\ref{fig:Ekin_distributions_20000protons_3D_smalleta10-9_uniform_thermalbath}) remains (nearly) uniform till $t = 6 t_S$ and afterward it is peaked around $\alpha = 0$ but at least an order of magnitude smaller than in all other proton cases.
\begin{figure*}
  \centering
	\subfloat{\includegraphics[width=2.3\columnwidth, trim= 3cm 0cm 0cm 0cm, clip=true]{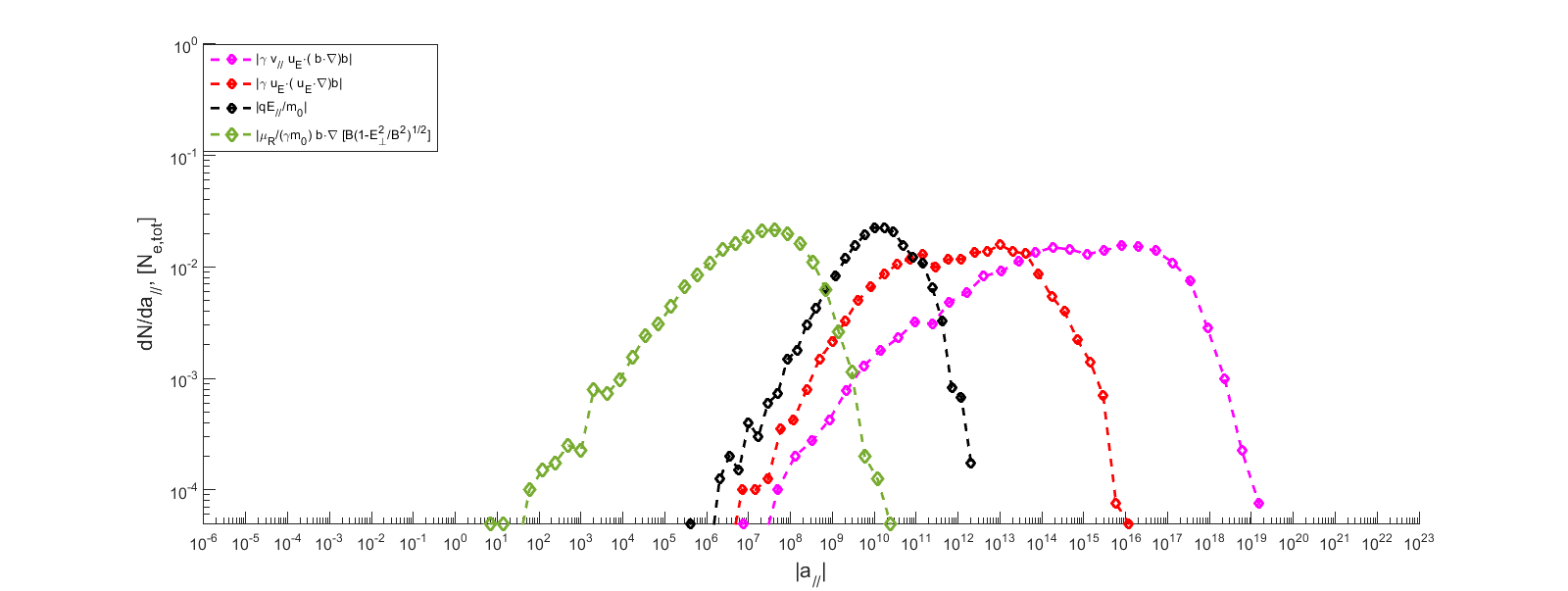}}
\caption{The number distribution of the four acceleration terms in the right-hand-side of equation (\ref{eq:gcastatic2}) for 20.000 electrons in case G3De at $t=9 t_S$. Compare to Fig.~\ref{fig:momentum_terms_t6} for cases A3De and C3De. The length of the flux rope is $6L$ and resistivity is set as $\eta_p = 10^{-5} \times \eta_{MHD} = 10^{-9}$. The magnetic mirror term (fourth term in equation \ref{eq:gcastatic2}) is indicated by a green line, the resistive electric field acceleration (third term in equation \ref{eq:gcastatic2}) by a black line and the two terms attributed to the change of direction of the magnetic field in red (polarization term in equation \ref{eq:gcastatic2}) and in magenta (magnetic curvature term in equation \ref{eq:gcastatic2}).}
\label{fig:momentum_terms_t9_both}
\end{figure*}
In Fig.~\ref{fig:momentum_terms_t9_both} contributions of the four separate terms in the momentum equation (\ref{eq:gcastatic2}) are quantified. Comparing to Fig.~\ref{fig:momentum_terms_t6} for cases A3De and C3De the effect of a finite flux rope length and low resistivity combined is shown for the four acceleration mechanisms. We see that at $t=9$ the magnetic curvature is the dominant acceleration mechanism. There is no peak in the distribution. The magnetic curvature acceleration is three orders of magnitude lower than in case C3De (with $\eta_p = 10^{-9}$ and periodic flux ropes) and one order of magnitude lower than in case A3De (with finite flux rope length and $\eta_p = 10^{-4}$). The other three acceleration mechanisms are at least three orders of magnitude smaller than the curvature acceleration. The curvature acceleration term in the GCA momentum equation (\ref{eq:gcastatic2}) is proportional to the parallel velocity of the particle. A particle following a (curved) field line accelerates parallel to the field line. Even in the initially straight current channel the field lines are curved due to the initial magnetic field distribution in equation (\ref{eq:eqB}).
Conclusively, a realistic magnetic Reynolds number and finite flux rope length result in a realistic maximum energy reached, power law index of the high energy tail of the kinetic energy distribution and the shape of the pitch angle distributions. Case G3De for electrons and case F3Dp for protons give the most realistic results for particle acceleration due to interacting flux ropes in the solar corona. In the next section we explore the effect of the finite length of a flux rope for the settings of case G3De.

\subsection{Effect of the length of the flux ropes on particle distributions}
\begin{figure*}
  \centering
\subfloat{\includegraphics[width=\columnwidth, clip=true]{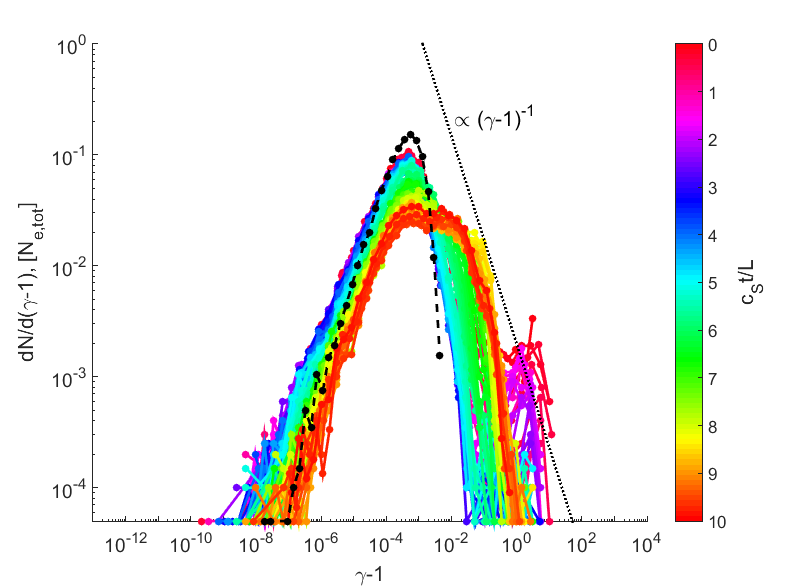}}   
 \subfloat{\includegraphics[width=\columnwidth, clip=true]{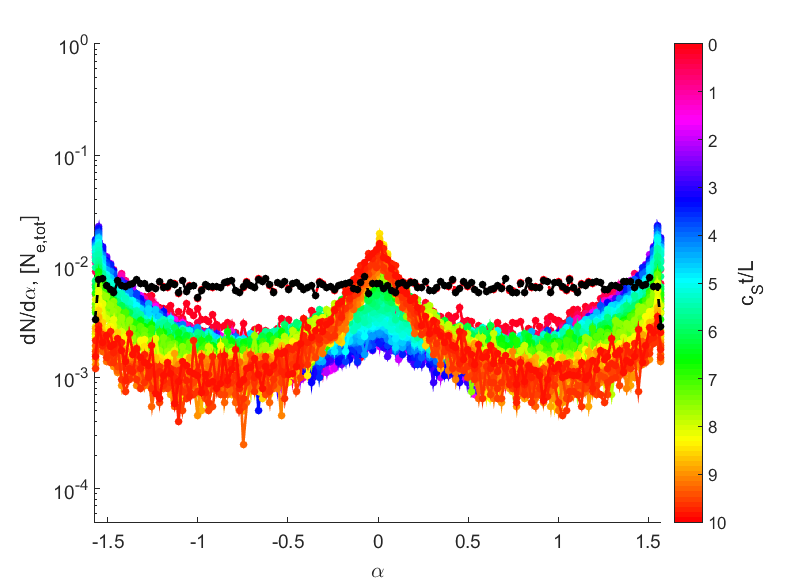}}
\caption{Distributions for 20000 electrons in case H3De with lowered resistivity $\eta_p = 10^{-9}$, initial thermal speed $v_{th,e}=\sqrt{2 k_B T \rho_0/(m_{p} p_0)}\sqrt{m_{p}/m_{e}}$ and now double the length $12L$ of the current channels (i.e. thermal bath applied). Compare to case G3De (Fig.~\ref{fig:Ekin_distributions_20000electrons_3D_smalleta10-9_uniform_massratio}) with half the length ($6L$) of the current channels. As a guide for the eye a typical power law distribution of the form $dN/d(\gamma-1) \propto (\gamma-1)^{-1}$ is plotted (dotted black line), corresponding to equal energy content in each decade of $\gamma-1$.}
\label{fig:Ekin_distributions_doublelength}
\end{figure*}

In case of periodic boundary conditions particles are either expelled from the flux rope or they travel through the flux rope until the simulation ends. When they are expelled from the flux rope they thermalize in the ambient medium until they leave the domain or they are caught into the flux ropes again. The particles causing the high energy peak are in the current channels for a long time and typically cycle through the current channel many times. These fast electrons traveling in a flux rope typically accelerate up to $\gamma_{max} \approx 1 \cdot 10^3$ if the flux rope is infinitely long (i.e. if no thermal bath is applied as in case C3De for electrons). During the simulation an electron with a parallel velocity $v_{\|} \approx c$ travels a maximum distance of $\Delta l \approx c \times 10 t_S \approx 2.6 \cdot 10^{13} $ cm, or over 4000 cycles through the flux ropes, if it is not expelled at some point. This is equal to $\Delta l\approx 3.6 \cdot 10^2$ solar radii in 10 soundcrossing times $t_S$. The medium energy tail in case C3De (see Fig.~\ref{fig:Ekin_distributions_20000electrons_3D_smalleta10-9}) consists of particles with an average Lorentz factor of $\gamma - 1 \approx 10^{-2}$. This is equal to a parallel velocity of $v_{\|} \approx 0.14 c$. Even these particles can travel a maximum distance of $\Delta l \approx 0.14c \times 10 t_S \approx 3.4 \cdot 10^{12} $ cm, or 52 solar radii through the flux ropes. Thermal bath boundary conditions limit this to one cycle, or $\Delta l = 6L \approx 8.6 \cdot 10^{-2}$ solar radii. To analyze how the length of the flux rope affects the energy distribution, case H3De has settings similar to case G3De, with the total length of the current channels $12L$ (or $1.7 \cdot 10^{-1}$ solar radii) instead of $6L$. All particles are initially in the current channel area. To obtain equal accuracy the resolution is also doubled in the $z$-direction. The shape of the distributions for case H3De in Fig.~\ref{fig:Ekin_distributions_doublelength} is similar to case G3De (Fig.~\ref{fig:Ekin_distributions_20000electrons_3D_smalleta10-9_uniform_massratio}). The maximum energy reached by the particles in the flux ropes with length $12L$ is approximately doubled compared to case G3De ($\gamma_{max} - 1 \lesssim 2 \cdot 10^1$ or $\mathcal{E} \lesssim 11$ MeV). Initially few particles accelerate in the flux tubes and the energy distribution in the left-hand panel remains largely Maxwellian. In the nonlinear phase, after $t = 6 t_S$ a non-thermal power-law tail forms due to the tilt-kink instability and subsequent reconnection. The pitch angle distribution in the right-hand panel remains nearly flat until $t = 6 t_S$, after which the peak around $\alpha = 0$ becomes dominant due to particles accelerating along the magnetic field.

\section{Conclusions}
\label{sect:conclusions}
We find that reconnection in the low plasma-$\beta$ regime drives efficient energy conversion from magnetic energy to kinetic energy and that electrons and protons are efficiently accelerated to non-thermal energy distributions. We observe two populations of high energy particles; A high energy peak of particles trapped inside the current channels at early times, where they accelerate efficiently along the magnetic field. Electrons reach maximum energies between $\sim$10 and $\sim$500 MeV. The maximum energy depends strongly on initial velocity distribution, plasma resistivity and the length of the flux tubes. A second population consists of electrons accelerating in the reconnection zones at late times in the nonlinear phase. These electrons generate a high energy tail in between the peak and the Maxwellian part of the distribution due to particles accelerating, in quite a narrow range between $\sim$0.5 MeV and $\sim$10 MeV. Protons reach maximum energy of approximately 1000 MeV in all cases.

We proposed two solutions to limit indefinite particle acceleration due to infinitely long flux ropes as found in \cite{Ripperda} for 2.5D simulations of low plasma-$\beta$ environments and extended this setup to full 3D simulations for electrons and protons. One solution is to apply a thermal bath along the length-direction of the flux ropes at the periodic boundaries (case A3De for electrons and B3Dp for protons). This solution effectively limits the length of the flux ropes to realistic scales and destroys fast particles leaving the domain. For every particle ending up in the thermal bath, a new thermal particle is injected at the opposite periodic boundary. This assures to keep a steady total number of particles. The particles maximum kinetic energy grows linearly with the length of the flux ropes and the time spent in the flux rope. In a flux rope of length $6L$ the maximum particle energy is limited to $\gamma - 1 \lesssim 100$, or $\mathcal{E} \lesssim 50$ MeV for electrons and $\gamma - 1 \lesssim 2 \cdot 10^{-2}$, or $\mathcal{E} \lesssim 960$ MeV for protons. The majority of particles reach a non-thermal energy before they reach the end of the flux rope. A high energy tail forms in all cases with limited flux rope length. The shape of the high energy tail and maximum energy depend on the initial conditions and resistivity set for the simulation.

The second solution counteracts acceleration in the direction of the resistive electric field, parallel to the magnetic field, as proposed by \cite{Zhou2}. A particle resistivity is applied that is typically lower than the MHD resistivity and results in realistic magnetic Reynolds number for solar corona conditions (case C3De for electrons and D3Dp for protons). This particle resistivity effectively lowers the electric field that is felt by the particles, but does not affect the MHD evolution and the formation of strong and thin current sheets. It limits the number of particles that accelerate to high energy. Despite having less particles in the high energy peak, the maximum particle energy is not limited due to the infinite length of the (periodic) flux ropes. Electrons accelerate to $\gamma - 1 \lesssim 1000$ or $\mathcal{E} \lesssim 500$ MeV and protons to $\gamma - 1 \lesssim 2 \cdot 10^{-2}$, or $\mathcal{E} \lesssim 960$ MeV. A high energy distribution forms in all cases and its shape and maximum energy depends on the initial conditions and the resistivity set.

Despite limited maximum particle energy an inverted power law index ($p < 0$) is found for the non-thermal distribution in the cases with low resistivity or limited flux rope length. This is due to the high energy peak developing from particles accelerating in the current channels at early times. These high energy particles dominate the energy census at late times. To limit the number of particles in the high energy tail we combined limiting the flux rope length and the resistivity to realistic values (cases E3De, G3De and H3De for electrons and F3Dp for protons). This results in setups with realistic magnetic Reynolds number and a flux rope length limited to a fraction of a solar radius. The maximum energy grows with the length of the flux ropes and for electrons $\mathcal{E} \lesssim 4$ MeV for length $60$ Mm and $\mathcal{E} \lesssim 11$ MeV for length $120$ Mm and for protons $\mathcal{E} \lesssim 1$ GeV. The shape of the slope of the high energy tail formed depends on the initial velocity distribution. Assuming energy equipartition between electrons and protons we find high energy power-law distributions $f(\mathcal{E}) \sim (\mathcal{E})^{-p}$ with  $p \geq 1$, as opposed to the inverted spectra found if the flux rope length is not limited and the resistivity is not lowered to realistic solar corona values. The high energy particles are not dominant in number, nor in energy, such that the test particle approximation is valid. In all these cases a part of the particle ensemble and its energy content is in the non-thermal distribution. These findings are in good agreement with the 2D kinetic PIC results of \cite{Li} for particle acceleration in force-free current sheets in low-$\beta$ electron-proton plasmas. The non-thermal particle distributions found can explain the efficient electron acceleration in low plasma-$\beta$ environments such as solar flares.

Without applying the two solutions proposed the whole ensemble of particles contains so much energy that kinetic feedback of the particles to the electromagnetic fields cannot be neglected. The assumption that the energy content of the particles is much lower than that of the fluid is invalid. A kinetic description allows particles to lose their energy to the fields through kinetic instabilities and particle-field interaction that is ignored in the test particle approximation. However at solar length scales, considering that reconnection occurs globally in the simulation box, PIC simulations are extremely demanding numerically. In cases with realistic magnetic Reynolds numbers and solar length scales for the flux ropes the number of accelerated particles is small compared to the total number of particles. The fields induced by these particles should not substantially affect the MHD evolution and reconnection. 

We applied a guiding centre approximation, ignoring the gyration of particles and reducing computing time. Via the guiding centre approximation we demonstrated that magnetic curvature is the leading acceleration mechanism in all cases. The magnetic field inside the current channels is curved initially and the kink instability introduces more curvature, making the curvature acceleration dominant at all times. The curvature acceleration is proportional to the velocity of a particle parallel to the magnetic field, enhancing the effect of the curvature. Particles are mainly accelerated parallel to the magnetic field and hence the resistive electric field, resulting in pitch angle distributions dominated by a peak at $\alpha = 0$ in all cases. The width and height of the pitch angle peak depends on the maximum parallel particle velocity and hence on particle energy. The $\alpha = 0$ peak shows an electron/proton asymmetry, previously observed by \cite{Gordovskyy}, caused by protons predominantly moving with positive parallel velocity and electrons mostly with negative parallel velocity along the magnetic field.

The guiding centre approximation is valid since the gyration radius remains much smaller than typical cell size in all runs. For protons, the gyration might not be negligible due to a larger mass and hence a larger gyroradius. To monitor the validity of our approach, results are compared to a run where proton gyration is not neglected and the full equations of motion (\ref{eq:lorentztens}) are solved in Appendix \ref{sect:appgyration} (case K3Dp). Applying the GCA has little to no effect on the energy distribution, compared to a case where particle gyration is fully resolved. The GCA is accurate in magnetically dominated, Newtonian plasmas under solar corona conditions. Results are also confirmed for both an ensemble of 200.000 particles (Appendix \ref{sect:appnumber}, case I3De) and for an initially uniform spatial distribution (Appendix \ref{sect:appspatial}, case J3De) to show that obtained statistics are accurate. Since kinetic feedback of the particles on the fields is neglected the accumulated current from energetic particles is unimportant and it is not necessary to increase the total number of particles. 

\section*{Acknowledgements}
This research was supported by projects GOA/2015-014 (2014-2018 KU Leuven) and the Interuniversity Attraction Poles Programme by the Belgian Science Policy Office (IAP P7/08 CHARM). OP is supported by the ERC synergy grant `BlackHoleCam: Imaging the Event Horizon of Black Holes' (Grant No. 610058).
The computational resources and services used in this work were provided by the VSC (Flemish Supercomputer Center), funded by the Research Foundation Flanders (FWO) and the Flemish Government - department EWI.
BR likes to thank Lorenzo Sironi, Fabio Bacchini, Norbert Magyar, Jannis Teunissen, Kirit Makwana, Matthieu Leroy and Dimitris Millas for fruitful discussions and comments.
%
%
%
%
\bibliographystyle{mnras}
\bibliography{mylib2} 
%
%
%
%
%
%
%
%
\appendix
\section{Validation tests}
\subsection{Number of particles}
\label{sect:appnumber}

\begin{figure}
  \centering
\subfloat{\includegraphics[width=\columnwidth, clip=true]{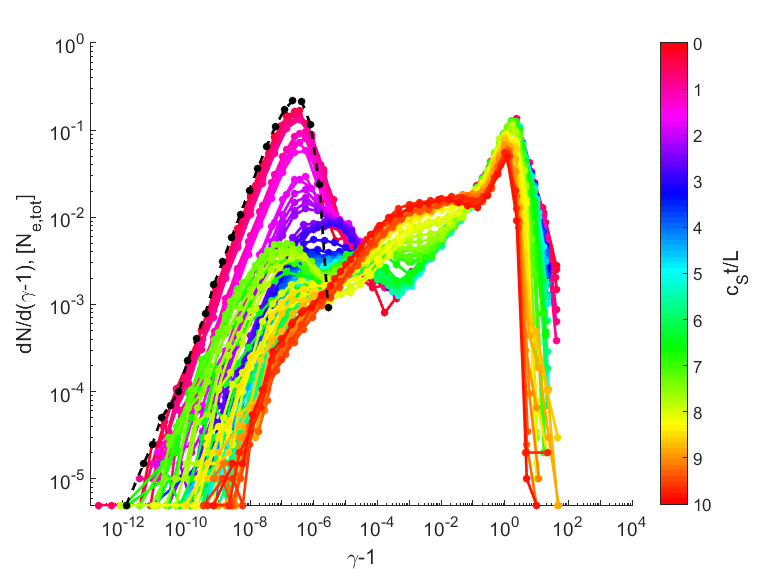}}
\caption{Kinetic energy distribution as in Fig.~\ref{fig:Ekin_distributions_20000electrons_3D} for case I3De with 200.000 electrons in 3D MHD with $\eta_p = 10^{-4}$ and $6L$ for the length of the current channels.}
\label{fig:Ekin_distributions_200000electrons_3D}
\end{figure}

To assure that the results are statistically accurate, a setup similar to case A3De is evolved with an initial distribution of 200.000 electrons for case I3De. Minor differences are visible in Fig.~\ref{fig:Ekin_distributions_200000electrons_3D}, mainly due to smoother distributions. The energy range and global features of both the kinetic energy distributions and the pitch angle distribution are in agreement with results for 20.000 electrons. Evolving more particles has no effect on the physical results, other than improving statistics, since there is no feedback of the particles on the fields, nor any particle-particle interaction.

\subsection{Spatial distribution}
\label{sect:appspatial}

\begin{figure}
  \centering
\subfloat{\includegraphics[width=\columnwidth, clip=true]{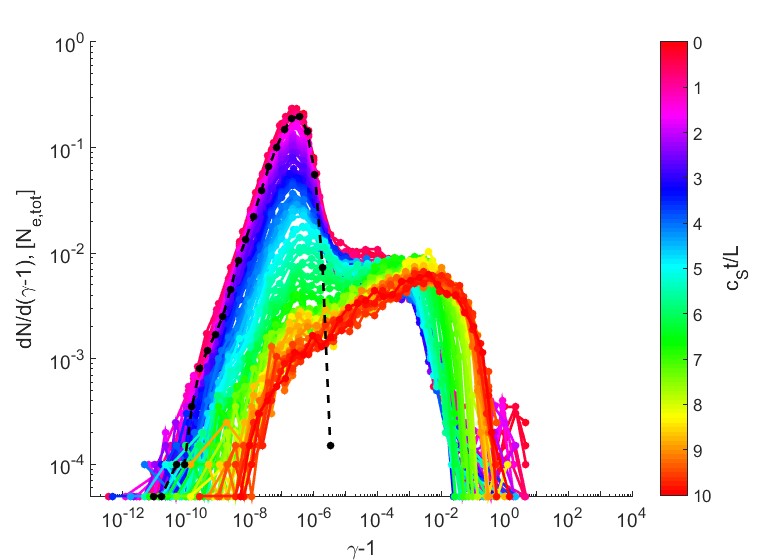}}
\caption{Kinetic energy distribution as in Fig.~\ref{fig:Ekin_distributions_20000electrons_3D_smalleta10-9_thermalisation} for case J3De with 20.000 electrons in 3D MHD, uniformly distributed in the domain with $\eta_p = 10^{-9}$ and $6L$ for the length of the current channels.}
\label{fig:Ekin_distributions_20000electrons_3D_smalleta10-9_uniform}
\end{figure}

Based on the findings of \cite{Ripperda} for 2.5D configurations, a fraction of $0.99$ of the particles is initially distributed in the area of the current channels $ -1L \leq x \leq 1L$; $-2L \leq y \leq 2L$; $ -3L \leq z \leq 3L$. In this way using 20.000 particles results in accurate statistics, similar to a run with 200.000 particles, and computational resources are mainly used on particles that are likely to accelerate. In 3D configurations a larger area of the box is filled with reconnecting magnetic field due to the kink instability. To monitor whether it is accurate to distribute most particles in the region of the current channels in 3D configurations a setup similar to case E3De with thermal bath and $\eta_p=10^{-9}$ is ran for case J3De, now with an initially uniform spatial distribution. Electrons are randomly and uniformly initialized in the whole box $ -3L \leq x \leq 3L$; $-3L \leq y \leq 3L$; $ -3L \leq z \leq 3L$. The resulting energy distributions in Fig.~\ref{fig:Ekin_distributions_20000electrons_3D_smalleta10-9_uniform} show the same shape and trend and the maximum energy $\gamma-1 \lesssim 5$ is equal to the maximum energy obtained in case E3De in Fig.~\ref{fig:Ekin_distributions_20000electrons_3D_smalleta10-9_thermalisation}. The main difference is that there are less particles in the high energy tail. However, there are not more particles remaining in the thermal distribution; The particles in he ambient medium, outside the current channels, leave the domain guided by the field lines on the timescales considered. Also the pitch angle is still peaked around $\alpha = 0$, but the peak is an order of magnitude lower than in case E3De. Distributing particles in the area of the current channels rather than in the ambient medium improves statistics on high-energy particles, but does not alter the physical features of the obtained spectra. This is in accordance with the 2.5D results of \cite{Ripperda}.

\subsection{Gyration}
\label{sect:appgyration}
\begin{figure}
  \centering
\subfloat{\includegraphics[width=\columnwidth, clip=true]{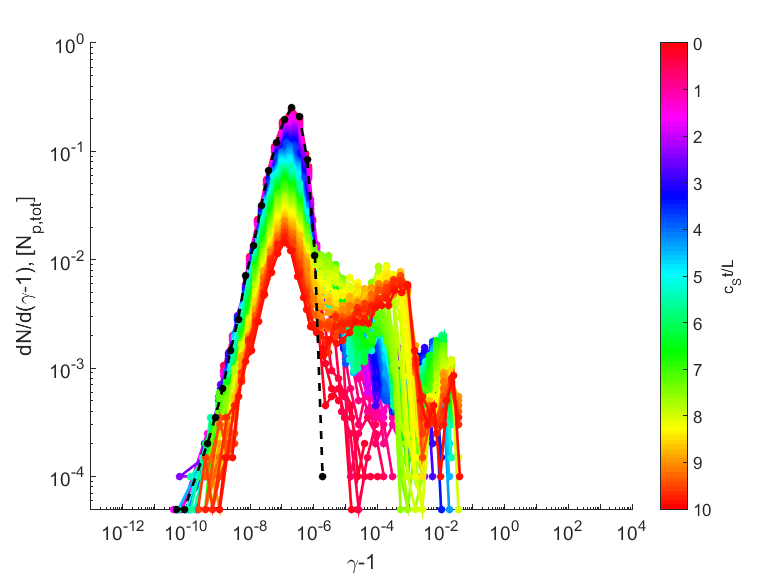}}   
\caption{Kinetic energy distribution as in Fig.~\ref{fig:Ekin_distributions_20000protons_3D_smalleta10-9_uniform} for case K3Dp with 20.000 protons in 3D MHD, solved with the full equations of motion, with $\eta_p = 10^{-9}$ and infinite length of the current channels.}
\label{fig:Ekin_distributions_20000protons_3D_smalleta10-9_uniform_Lorentz}
\end{figure}

Protons have a larger gyroradius than electrons due to their mass and therefore the GCA is most likely to break down in proton simulations. To monitor the validity of the GCA, the full particle equations of motion (\ref{eq:lorentztens}) including gyromotion are evolved with a Boris scheme for protons in the same MHD background. In Fig.~\ref{fig:Ekin_distributions_20000protons_3D_smalleta10-9_uniform_Lorentz} we show the kinetic energy distribution for 20000 protons in a setup where particle gyration is resolved. Case D3Dp is chosen for comparison to confirm that the high energy peak found is not a numerical artifact. The energy distribution shows a very similar shape. A peak develops due to particles accelerating in the current channels and the maximum energy reached is limited by the length of the current channels and the thermal bath boundary conditions, such that the maximum energy is similar to case D3Dp, with $\gamma - 1 \leq 4 \cdot 10^{-2}$ ($\mathcal{E} \lesssim 976$ MeV). The high energy tail shows an inverted power law index, developing at similar time as in case D3Dp. Besides minor differences in the thermal part of the spectrum, due to the random assignment of a thermal velocity component $v_x$, $v_y$ and $v_z$ rather than $v_{\|}$ and $v_{\perp}$ in the GCA setups, the global aspects are similar to the GCA results in Fig~\ref{fig:Ekin_distributions_20000protons_3D_smalleta10-9_uniform}. For electrons the differences are expected to be even smaller, due to a smaller mass and hence completely negligible gyroradius.

\section{Erratum: Reconnection and particle acceleration in interacting flux ropes -- II. 3D effects on test particles in magnetically dominated plasmas}

The paper "Reconnection and particle acceleration in interacting flux ropes -- II. 3D effects on test particles in magnetically dominated plasmas" was published in MNRAS, Volume 471, Issue 3 (2017). After the publication of the article an error was found in a postprocessing script affecting several figures. The numerical data and the results of the calculations are correct and the conclusions drawn are too. Nevertheless, the magnitude of the curvature acceleration term ($|\gamma\mathbf{u_E}\cdot \left(v_{\|}\left(\mathbf{\hat{b}}\cdot\nabla\right)\mathbf{\hat{b}}\right)|$) and the magnitude of the polarization acceleration term ($|m_0\gamma\mathbf{u_E}\cdot \left(\left(\mathbf{u_E}\cdot\nabla\right)\mathbf{\hat{b}}\right)|$), were taken too large by a factor $c$ (speed of light) while manufacturing Figures 7, 12, 13 and 17 in the original manuscript. This affects the scale of the bottom panels in Figure 7 and the right-hand panel in Figure 12 in the original manuscript, where the colour bar ticks should be divided by the speed of light $c$ in CGS units. The conclusions drawn from these figures remain fully valid since only the scale has changed.

Removing the erroneous factor $c$ in the curvature acceleration term and the polarization acceleration term in the postprocessing script also affects Figures 13 and 17 where the number distributions of the curvature acceleration and the polarization acceleration (magenta and red curves in the original Figures respectively) are shifted to the right by a factor $c$. The conclusion that curvature acceleration is dominant at all times is therefore incorrect and should be replaced by the conclusion that the curvature acceleration is the second most dominant mechanism of particle acceleration in cases A3De and C3De (Fig.~\ref{fig:momentum_terms_t6}) and that it grows strongly in the nonlinear regime after $t \approx 6t_S$ in case C3De with a lowered resistivity. In case G3De (Fig.~\ref{fig:momentum_terms_t9_both}) the parallel acceleration is limited by both a lowered resistivity and a limited length of the flux rope, resulting in a less dominant curvature acceleration and a more dominant mirror mechanism (green curve). The resistive electric field acceleration is dominant as was already remarked in the conclusions of the original manuscript, with the remark that this term depends linearly on the resistivity that is arbitrarily set. Therefore the approach to lower the particle resistivity to a value realistic for the solar corona is still justified by the correction made here. The general conclusions in the final section of the original manuscript are unaffected by this correction.

\begin{figure*}
  \centering
	\subfloat{\includegraphics[width=2.3\columnwidth, trim= 3cm 0cm 0cm 0cm, clip=true]{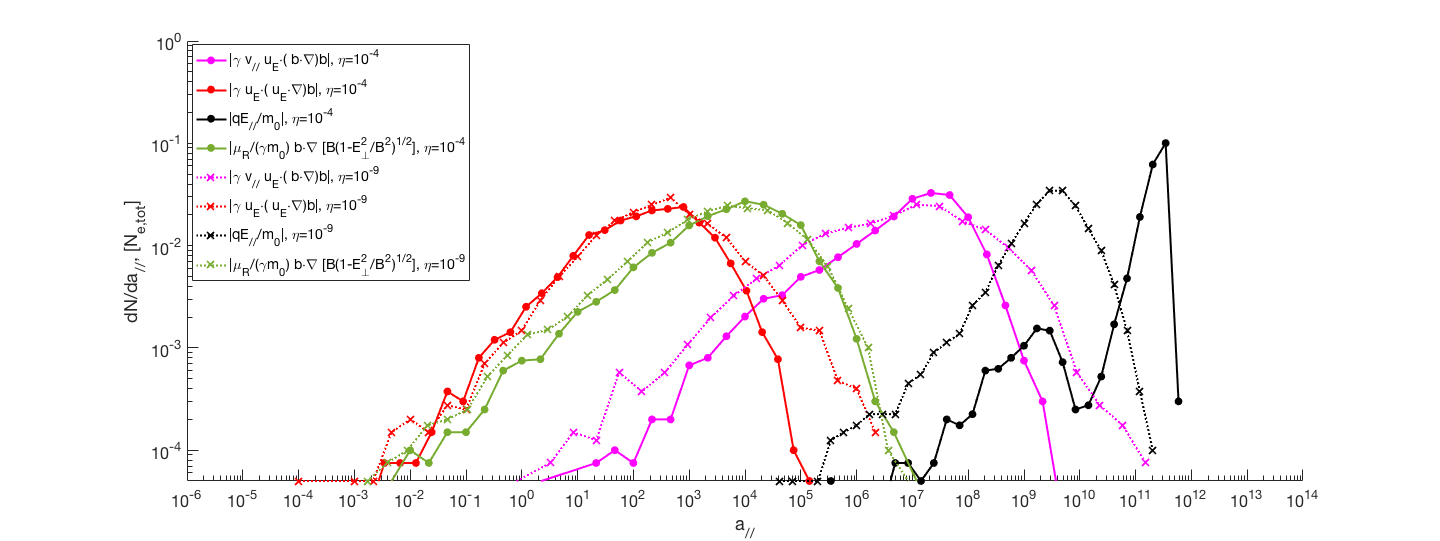}}
	
	\subfloat{\includegraphics[width=2.3\columnwidth, trim= 3cm 0cm 0cm 0cm, clip=true]{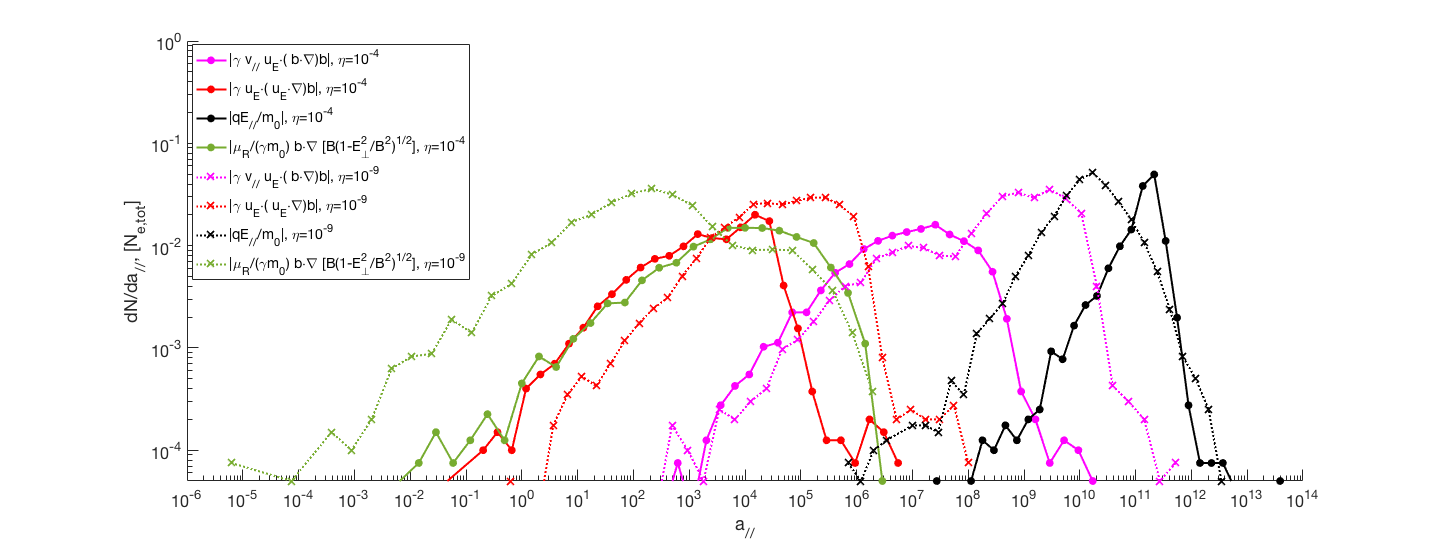}}
\caption{Comparison of the number distribution of the four acceleration terms in the right-hand-side of equation (6) in the original manuscript for 20.000 electrons in case C3De (dotted lines with crosses as indicators) and case A3De (solid lines with dots as indicators) before nonlinear phase at $t=6 t_S$ (top panel) and during nonlinear phase at $t=9 t_S$ (bottom panel). In case C3De periodic boundary conditions are applied (i.e. infinitely long flux ropes) and $\eta_p = 10^{-5} \times \eta_{MHD} = 10^{-9}$ and in case A3De the flux rope length is $6L$ and resistivity is set $\eta_p = \eta_{MHD} = 10^{-4}$. The magnetic mirror term (fourth term in equation 6) is indicated by a green line, the resistive electric field acceleration (third term in equation 6) by a black line and the two terms attributed to the change of direction of the magnetic field in red (polarization term, the second term in equation 6) and in magenta (magnetic curvature term, the first term in equation 6).}
\label{fig:momentum_terms_t6}
\end{figure*}

\begin{figure*}
  \centering
	\subfloat{\includegraphics[width=2.3\columnwidth, trim= 3cm 0cm 0cm 0cm, clip=true]{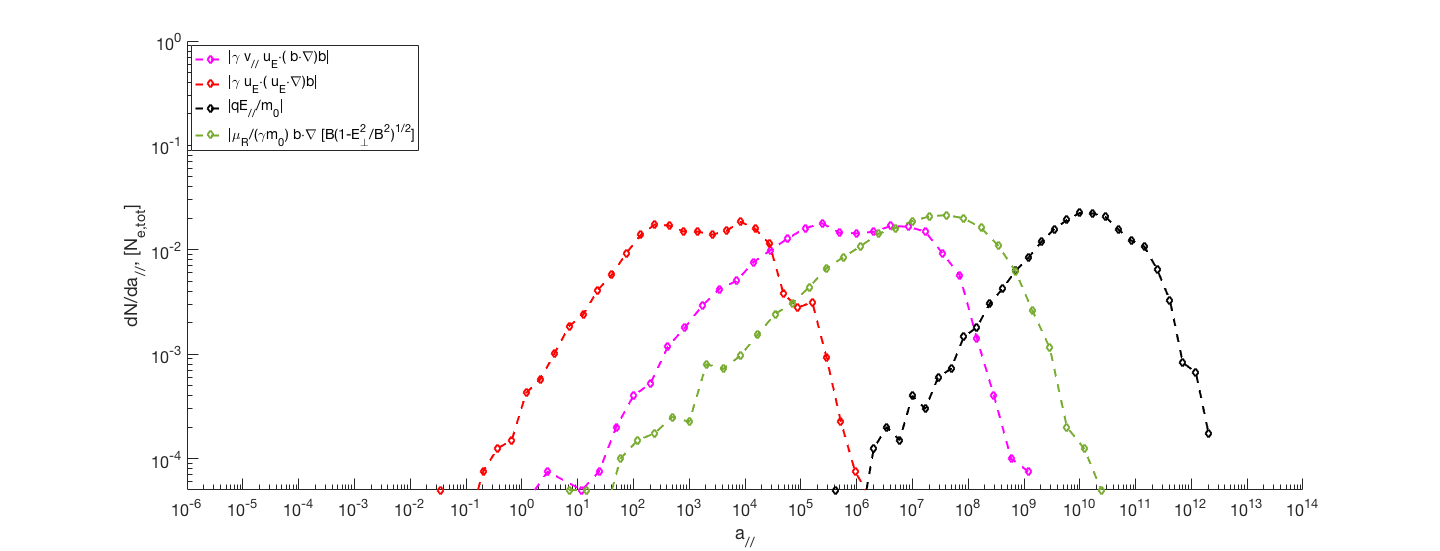}}
\caption{The number distribution of the four acceleration terms in the right-hand-side of equation (6) in the original manuscript, for 20.000 electrons in case G3De at $t=9 t_S$. Compare to Fig.~\ref{fig:momentum_terms_t6} for cases A3De and C3De. The length of the flux rope is $6L$ and resistivity is set as $\eta_p = 10^{-5} \times \eta_{MHD} = 10^{-9}$. The magnetic mirror term (fourth term in equation 6) is indicated by a green line, the resistive electric field acceleration (third term in equation 6) by a black line and the two terms attributed to the change of direction of the magnetic field in red (polarization term in equation 6) and in magenta (magnetic curvature term in equation 6).}
\label{fig:momentum_terms_t9_both}
\end{figure*}
%
%
%
\bsp	
\label{lastpage}
\end{document}